\documentclass{beamer}

\usepackage{amssymb,amsmath}

\usepackage{amsmath,amssymb,amsfonts,epsfig}
\usepackage[active]{srcltx}
\usepackage{graphics}
\usepackage{amssymb,amsmath}
\usepackage{cite}
\usepackage{color}
\usepackage[usenames,dvipsnames]{pstricks}
\usepackage{pstricks}
\usepackage{epsfig}
\usepackage{pst-grad} 
\usepackage{pst-plot} 

\mode<presentation>{\usetheme{Madrid}}

\title[Speaker: L.V.~Laperashvili]{$E_6$ Unification and the Hidden 
Sector of the Universe}

\author[ITEP, Moscow]{\small \textbf{C.R.~Das} \\
Centre for Theoretical Particle Physics, Technical University of
Lisbon, \\ Avenida Rovisco Pais, 1 1049-001 Lisbon,
Portugal\\[5mm]
\textbf{\alert{L.V.~Laperashvili}} \\
The Institute of Theoretical and Experimental Physics,\\
Bolshaya Cheremushkinskaya, 25, 117218 Moscow, Russia\\[5mm]
\textbf{H.B.~Nielsen} \\Theoretical Particle Physics and Cosmology\\
Niels Bohr Institute, Blegdamsvej 17, 2100 Copenhagen, Denmark\\[5mm]
\textbf{A.~Tureanu} \\
Department of Physics, University of Helsinki\\ and Helsinki
Institute of Physics, P.O.Box 64, FIN-00014 Helsinki, Finland}

\date{\today}

\AtBeginSection[] 
{\begin{frame}<beamer> 
\frametitle{Outline}
\tableofcontents[currentsection]
\end{frame}}

\newcommand{\lb}{\label}

\newcommand{\bc}{\begin{center}}
\newcommand{\ec}{\end{center}}
\newcommand{\bd}{\begin{displaymath}}
\newcommand{\ed}{\end{displaymath}}
\newcommand{\be}{\begin{equation}}
\newcommand{\ee}{\end{equation}}
\newcommand{\ba}{\begin{array}}
\newcommand{\ea}{\end{array}}
\newcommand{\bea}{\begin{eqnarray*}}
\newcommand{\eea}{\end{eqnarray*}}
\newcommand{\bt}{\begin{tabular}}
\newcommand{\et}{\end{tabular}}
\newcommand{\un}{\underline}

\newcommand{\bp}{\begin{picture}}
\newcommand{\ep}{\end{picture}}
\newcommand{\bfi}{\begin{figure}}
\newcommand{\efi}{\end{figure}}

\begin{document}

\frame{\titlepage}

\section*{Outline}
\begin{frame}
\frametitle{Outline}
\tableofcontents
\end{frame}

\begin{frame}
\noindent
The present paper is devoted to the new Cosmological Model with
$E_6$ unification at the early stage of the Universe, explaining
why cosmological constant is extremely small.
\vskip 0.2in
\noindent
The ideas of this talk were published in the following references:

\begin{enumerate}
\item
C.R. Das, L.V. Laperashvili, H.B. Nielsen, A. Tureanu,\\
{\bf Mirror World and Superstring-Inspired Hidden Sector of the
Universe, Dark Matter and Dark Energy};\\
arXiv:1101.4558 [hep-ph], to appear in Phys.Rev.D. (2011).
\item
C.R. Das, L.V. Laperashvili,  H.B. Nielsen, A. Tureanu,\\
{\bf Baryogenesis in Cosmological Model with Superstring-Inspired
$E_6$ Unification};\\
Phys.Lett.B {\bf 696} (2011), 138; arXiv:1010.2744 [hep-ph].
\item C.R. Das, L.V. Laperashvili, A. Tureanu, \\
Eur.Phys.J.C {\bf 66} (2010), 307; arXiv:0902.4874 [hep-ph].
\end{enumerate}

\end{frame}

\begin{frame}
\noindent
They also were presented at the following conferences:

\begin{enumerate}
\item C.R. Das, L.V. Laperashvili, A. Tureanu, \\ INVISIBLE
UNIVERSE INTERNATIONAL CONFERENCE: {\it Toward a new cosmological
paradigm}, Paris, France, 29 Jun - 3 Jul 2009.\\
AIP Conf.Proc. {\bf 1241} (2010), 639; arXiv:0910.1669 [hep-ph].
\item C.R. Das, L.V. Laperashvili, A. Tureanu, {\bf Superstring-Inspired $E_6$ Unification, 
Shadow Theta-Particles and Cosmology}\\ A talk given by Larisa Laperashvili at the 
International Bogolyubov Conference, Moscow-Dubna, August, 2009,\\
Phys.Part.Nucl. {\bf 41} (2010), 965; arXiv:1012.0624 [hep-ph].
\item C.R. Das and L.V. Laperashvili, {\bf $E_6$ unification and Cosmology}, in:
Proceedings of the XIII Int.Conference ``Selected Problems of Modern
Physics", dedicated to D.I. Blokhintsev, Dubna, August 2008.
\end{enumerate}

\end{frame}

\begin{frame}
\noindent
These papers are based on the previously published ideas:

\begin{block}{}
(from now: Refs~{\color{red}Hung et al. and Das-Laperashvili})
\end{block}

\end{frame}

\section{Introduction}

\begin{frame}
\frametitle{Introduction}
\noindent
Modern models for Dark Energy (DE) and Dark Matter (DM) are based
on precise measurements in cosmological and astrophysical
observations:

\end{frame}

\begin{frame}
\frametitle{Introduction}
\noindent
See also:

\end{frame}

\begin{frame}
\frametitle{Introduction}
\noindent
Supernovae observations by the Supernovae Legacy Survey (SNLS),
cosmic microwave background (CMB), cluster data and baryon
acoustic oscillations by the Sloan Digital Sky Survey (SDSS) fit
the equation of state for DE:

\begin{block}{}\centerline
{$w = p/\rho$}
\end{block}
\vskip 0.2in
\noindent
with constant $w$ and give the following result:

\begin{block}{}\centerline
{$w = -1.023 \pm 0.090 \pm 0.054.$}
\end{block}

\end{frame}

\begin{frame}
\frametitle{Introduction}
\noindent
The cosmological constant ($CC$), the vacuum energy
density of the Universe, also is given by the recent astrophysical
measurements:

\begin{block}{}\centerline
{$CC = \rho_{vac}\approx (2.3\times 10^{-3}\,\,{\rm eV})^4.$}
\end{block}
\vskip 0.2in
\noindent
The value $w=-1$ and a tiny cosmlogical constant are
consistent with the present quintessence model of accelerating
expansion of the Universe (so-called $\Lambda CDM$ scenario):

\end{frame}

\section{Superstring theory and $E_6$ Unification}

\begin{frame}
\frametitle{Superstring theory and $E_6$ Unification}
\noindent
Our model is based on the following assumptions:

\begin{itemize}
\item Grand Unified Theory is inspired by the
Superstring theory.
\end{itemize}
\vskip 0.2in
\noindent
The {\color{red}heterotic} superstring theory $E_8\times E'_8$ 
was suggested as a more realistic model for the
unification of all fundamental gauge interactions with gravity:

\end{frame}

\begin{frame}
\frametitle{Superstring theory and $E_6$ Unification}
\noindent
This ten-dimensional Yang-Mills theory can undergo  the
spontaneous breakdown of compactification. The integration over
six compactified dimensions of the $E_8$ superstring
theory leads to the effective theory with the $E_6$
unification in the four-dimensional space:

\begin{itemize}
\item Superstring theory predicts $E_6$ unification occurring
at the high energy scale $M_{E_6}\approx 10^{18}$ GeV
\end{itemize}
\end{frame}

\begin{frame}
\frametitle{$E_6$ Unification}
\noindent
Three 27-plets of $E_6$ contain three families of
quarks and leptons, including right-handed neutrinos.
\vskip 0.2in
\noindent
We omit generation subscripts, for simplification.
\vskip 0.2in
\noindent
Matter fields (quarks, leptons and scalar fields) of 27-plet
decompose under $SU(5)\times U(1)_X$ subgroup as follows:

\end{frame}

\begin{frame}
\frametitle{$E_6$ Unification}
\noindent
$$        27 \to (10,1)  + (\bar 5, 2)+
          (5,-2)+ (\bar 5,-3)  + (1,5) + (1,0).$$
\vskip 0.2in
\noindent
These representations decompose under the groups with the breaking
\vskip 0.2in
\noindent
$$SU(5)\times U(1)_X \to SU(3)_C\times SU(2)_L\times
U(1)_Z\times U(1)_X.$$
\vskip 0.2in
\noindent
We consider the following $U(1)_Z\times  U(1)_X$
charges of matter fields:
\vskip 0.2in
\noindent
$$Z=\sqrt{\frac{5}{3}}Q^Z,\,X=\sqrt{40}Q^X.$$

\end{frame}

\begin{frame}
\frametitle{$E_6$ Unification}
\noindent
We have the following assignments of particles:
\vskip 0.2in
\noindent
\bea
       (10,1) \to Q = &\left(\begin{array}{c}u\\
                                          d \end{array}\right) &\sim
                         \left(3,2,\frac 16,1\right),\nonumber\\
&u^{\rm\bf c} &\sim \left(\bar3,1,-\frac 23,1\right),\nonumber\\
&e^{\rm\bf c} &\sim \left(1,1,1,1\right).    \\
(\bar 5,2) \to &d^{\rm\bf c}&\sim \left(\bar 3,1,\frac
13,2\right),
\nonumber\\
L = &\left(\begin{array}{c}e\\
                                             \nu \end{array}\right) &\sim
                         \left(1,2,-\frac 12,2\right),              \\
(1,5) \to & S\,\,\, &\sim \,\,\left(1,1,0,5\right). \eea
\end{frame}

\begin{frame}
\frametitle{$E_6$ Unification}
\noindent
\bea
        (5,-2) \to& D&\sim \left(3,1,-\frac 13,-2\right),\nonumber\\
                   h = &\left(\begin{array}{c}h^+\\
                                               h^0 \end{array}\right) &\sim
                         \left(1,2,\frac 12,-2\right).
                                                             \\
    (\bar 5,-3) \to &D^{\rm\bf c} &\sim \left(\bar 3,1,\frac 13,-3\right),
\nonumber\\
                     h^{\rm\bf c} = &\left(\begin{array}{c}h^0\\
                                               h^- \end{array}\right) &\sim
                         \left(1,2,-\frac 12,-3\right).
\eea
\end{frame}

\begin{frame}
\frametitle{$E_6$ Unification}
\noindent
Also
$$(1,5) \to S \sim  (1,5,0,0),$$

the SM-singlet field S, which carries nonzero $U(1)_X$ charge.
\vskip 0.2in
\noindent
The light Higgs doublets are accompanied by the heavy colour
triplets of exotic quarks (or `diquarks') $D,D^{\rm\bf c}$ which
are absent in the SM.

\begin{block}{}
The right-handed heavy neutrino is a singlet field of $E_6$:
\vskip 0.2in
\noindent
$$(1,0) \to N^{\rm\bf c} \sim  (1,1,0,0).$$
\noindent
It is quite important for baryogenesis.
\end{block}

\end{frame}
\section{The breaking $E_6$ in O- and Sh-worlds}

\begin{frame}
\frametitle{The breaking $E_6$ in the O- and Sh-worlds}
\noindent
It is well known (see, for example:

\vskip 0.2in
\noindent
As we have seen, there are three ways of breaking the $E_6$ group:

\begin{enumerate}
\item[(i)] $E_6\to SU(3)_1\times SU(3)_2\times SU(3)_3$,
\item[(ii)] $E_6\to SO(10)\times U(1)$,
\item[(iii)] $E_6\to SU(6)\times SU(2)$.
\end{enumerate}

\end{frame}

\begin{frame}
\frametitle{The breaking $E_6$ in the O- and Sh-worlds}
\noindent
Previously we have considered the model with breaking

\begin{enumerate}
\item[(1)] $  E_6 \to SU(3)_C\times SU(3)_L\times SU(3)_R.$
\end{enumerate}
\vskip 0.2in
\noindent
Of course, such a Universe could exist, but then it is difficult
to explain a tiny value of $  CC$.

\begin{block}{}
In the present investigation we adopt for the O-world the breaking
         $  E_6\to SO(10)\times U(1),$
\end{block}

\begin{block}{}
while for the Sh-world we consider the breaking $ 
E'_6\to SU(6)'\times SU(2)'.$
\end{block}

\end{frame}

\section{Mirror and Shadow worlds}

\begin{frame}
\frametitle{Mirror and Shadow worlds}
\noindent
Superstring theory has led to the speculation that there may exist
in the Universe another form of matter - ``{\color{red}shadow matter}'':

\vskip 0.2in

\begin{block}{}
This shadow matter interacts with ordinary matter only
via gravity, or gravitational-strength interactions.
\vskip 0.2in

\alert{Shadow world} is an extension of the concept of the
\underline{Mirror World (MW)} --- a mirror duplication of our
Ordinary World (OW).
\end{block}

\end{frame}

\begin{frame}
\frametitle{Mirror and Shadow worlds}
\noindent
Lee and Yang were the first to suggest such a duplication of
the worlds, which restores the left-right symmetry of Nature:

\vskip 0.2in
\noindent
They introduced a concept of right-handed particles,
but their R-world was not hidden.

introduced the term `Mirror Matter' and `Mirror World' (MW) as
the hidden sector of our Universe. They demonstrated that
a mirror copy of the ordinary world interacts with our visible O-world
only via gravity or other very weak interactions.

\end{frame}

\begin{frame}
\frametitle{Mirror and Shadow worlds}
\noindent
The next assumptions:

\begin{itemize}
\item We assume that there exists a {\color{red}Shadow World (Sh-world)}
(hidden sector), parallel to our ordinary (visible) world, but it is
not identical with our O-world, having different symmetry
groups.
\item Shadow world describes the Dark Energy (DE) and Dark Matter
(DM) existing in our Universe:

see References {\color{red}Hung et al. and Das-Laperashvili}.
\item We assume that $  E_6$ unification had a place in
the Ordinary and Mirror worlds at the early stage of our Universe.
This means that at high energy scale $ \approx
10^{18}\,\,{\rm{GeV}}$  the Mirror World exists and the
group of symmetry of the Universe is $  E_6\times E'_6$

(the superscript `prime' denotes the M- or Sh-world).
\end{itemize}

\end{frame}

\section{The particle content in the ordinary and mirror worlds}

\begin{frame}
\frametitle{The particle content in the ordinary and mirror worlds}
\noindent
If we have in Nature ordinary and mirror worlds then at low
energies we can describe them by a minimal symmetry $$G_{SM}\times G'_{SM},$$
\par\noindent where $$G_{SM} = SU(3)_C\times SU(2)_L\times U(1)_Y$$ 
stands for the Standard Model ($SM$) of observable particles: three 
generations of quarks and leptons and the Higgs boson.
\par\noindent Then $$  G'_{SM} =
SU(3)'_C\times SU(2)'_L\times U(1)'_Y$$ is its mirror gauge
counterpart having three generations of mirror quarks and leptons
and the mirror Higgs boson.

\begin{alertblock}{}
The M-particles are singlets of $  G_{SM}$ and\\
the O-particles are singlets of $  G'_{SM}$.
\end{alertblock}

\end{frame}

\begin{frame}
\frametitle{The particle content in the ordinary and mirror worlds}

\begin{block}{}
Including the Higgs bosons $ \Phi$, we have the following
$  SM$ content of the O-world: 
\vskip 0.2in
\noindent
$ \rm{L-set}: \quad (u,d,e,\nu,\tilde u,\tilde d,\tilde
e,\tilde N)_L\,,$ $ \Phi_u,\,\Phi_d;$

$ \rm{\tilde R-set}: \quad (\tilde u,\tilde d,\tilde
e,\tilde \nu,u,d,e,N)_R\,,$ $ \tilde \Phi_u,\,\tilde
\Phi_d;$ 
\noindent
\vskip 0.2in
with the antiparticle fields: $ \tilde \Phi_ {u,d} =
\Phi^*_{u,d},\,\,$ $  \tilde \psi_R =
C\gamma_0\psi_L^*\,\,$ and $ \tilde \psi_L =
C\gamma_0\psi_R^*.$
\end{block}



\begin{block}{}
Considering the minimal symmetry $  G_{SM}\times
G'_{SM}$, we have the following particle content in the M-sector:
\vskip 0.2in
\noindent
$ \rm{ L'-set}: \quad (u',d',e',\nu',\tilde u',\tilde
d',\tilde e',\tilde N')_L\,,$ $ \Phi'_u,\,\Phi'_d ;$

$ \rm{\tilde R'-set}: \quad (\tilde u',\tilde d',\tilde
e',\tilde \nu',u',d',e',N')_R\,,$ $ \tilde
\Phi'_u,\,\tilde \Phi'_d.$
\end{block}

\end{frame}

\section{Mirror world with broken mirror parity}

\begin{frame}
\frametitle{Mirror world with broken mirror parity}
\noindent
If the ordinary and mirror worlds are identical, then O- and M-particles
should have the same cosmological densities. But this is in the immediate
conflict with recent astrophysical measurements.
\vskip 0.2in
\noindent
Mirror parity (MP) is not conserved, and the
ordinary and mirror worlds are not identical:

\noindent
From here \textcolor{blue}{Berezhiani-Dolgov-Mohapatra}.

\end{frame}

\begin{frame}
\frametitle{Mirror world with broken mirror parity}
\noindent
Similar ideas were considered in references:

\end{frame}

\begin{frame}
\frametitle{Mirror world with broken mirror parity}
\noindent
In the case of non-conserved MP the VEVs of the Higgs doublets
$  \Phi$ and $  \Phi'$ are not equal:  
$$\langle\Phi\rangle=v,\qquad \langle\Phi'\rangle=v',\qquad v\neq v'.$$
\vskip 0.2in
\noindent
We introduced the parameter
characterizing the violation of MP: $$  \zeta =
\frac{v'}{v} \gg 1. $$ 
\vskip 0.2in
\noindent Then the masses of fermions and massive
bosons in the mirror world are scaled up by the factor $
\zeta$ with respect to the masses of their counterparts in the
ordinary world: $$  m'_{q',l'} = \zeta m_{q,l}, \qquad
M'_{W',Z',\Phi'} = \zeta M_{W,Z,\Phi}, $$ 
\noindent while photons and gluons
remain massless in both worlds.
\end{frame}

\begin{frame}
\frametitle{Mirror world with broken mirror parity}
\noindent
The value of $ \zeta$ was estimated by astrophysical
implications in Refs.~ {\color{blue}Berezhiani-Dolgov-Mohapatra}:
$ \zeta\approx 30.$

\noindent
In the language of neutrino physics:

\begin{block}{}
the O-neutrinos $ \nu_e,\,\,\nu_{\mu},\,\,\nu_{\tau}$ are
{\color{red}active neutrinos},
\end{block}

\begin{block}{}
while the M-neutrinos $
\nu'_e,\,\,\nu'_{\mu},\,\,\nu'_{\tau}$ are {\color{red}sterile
neutrinos}.
\end{block}


\noindent
If MP is conserved ($ \zeta = 1$), then neutrinos of two
sectors are strongly mixed. But it seems that the situation with
the present experimental and cosmological limits on the
active-sterile neutrino mixing do not confirm this result. 
\vskip0.2in
\noindent
If instead MP is spontaneously broken, and $ \zeta >> 1$,
then the active-sterile mixing angles should be small:$
\theta_{\nu\nu'}\sim \frac 1{\zeta}. $ As a result, we have the
following relation between the masses of the light left-handed
neutrinos: $  m'_{\nu}\approx \zeta^2 m_{\nu}.$
\end{frame}

\section{Seesaw scale in the ordinary and mirror worlds}

\begin{frame}
\frametitle{Seesaw scale in the ordinary and mirror worlds}
\noindent
In the context of the SM, theory predicts that so called
right-handed neutrinos $  N_a$ with large Majorana mass
terms have equal masses in the O- and M(Sh)-worlds:

\begin{block}{}
$$ M'_{\nu,a} = M_{\nu,a}.$$
\end{block}
\noindent
Heavy right-handed neutrinos are created at seesaw scale $
 M_R$ (or $  M'_R$) in the O- (or M(Sh)-)world. 
\vskip 0.2in
\noindent
Theory predicts that even in the model with broken mirror parity, we
have the same seesaw scales in both worlds:

\begin{block}{}
       $$ M'_R = M_R. $$
\end{block}

\end{frame}

\section{The breaking $  E_6$ in the O- and Sh- worlds}

\begin{frame}
\frametitle{The breaking $  E_6$ in the O- and Sh- worlds}
\noindent
It is well known (see, for example:

\noindent
As was mentioned, three ways of breaking the $  E_6$ group:

\begin{enumerate}  
\item[(i)] $E_6 \to SU(3)_1\times SU(3)_2\times SU(3)_3,$
\item[(ii)] $E_6\to SO(10)\times U(1),$
\item[(iii)] $E_6\to SU(6)\times SU(2).$
\end{enumerate}
\noindent
The first case was considered in our paper:

\noindent
From here Ref. DLNT.
\end{frame}

\begin{frame}
\frametitle{The breaking $  E_6$ in the O- and Sh- worlds}
\noindent
Here we have investigated the possibility of the breaking:
\noindent
$$E_6\to SU(3)_C\times SU(3)_L\times SU(3)_R$$
\noindent
in both O- and M-worlds, with broken MP.
\vskip 0.2in
\noindent
The model has the merit of an attractive simplicity.
\vskip 0.2in
\noindent
However, in such a model we are unable to explain the tiny $CC$,
given by astrophysical measurements, because in this case we have
in both worlds the low-energy limit of the SM, which forbids a
large confinement radius (i.e. small energy scale) of any
interaction.

\begin{block}{}
It is quite impossible to obtain the same $  E_6$
unification in the O- and M-worlds with the same breakings $
  (ii)$ or $   (iii)$ in both worlds if  mirror
parity MP is broken.
\end{block}

\end{frame}

\begin{frame}
\frametitle{The breaking $  E_6$ in the O- and Sh- worlds}
\noindent
In this case, we are forced to assume different breakings of the
$  E_6$ unification in the O- and Sh-worlds:
\begin{block}{}
$$  E_6 \to SO(10)\times U(1) \qquad {\rm in\,\,\, O-world}, $$
$$   E'_6 \to SU(6)'\times SU(2)'\qquad {\rm in\,\,\, Sh-world}, $$
\end{block}
\noindent
This breaking explains the small value of the $CC$ by condensation
of fields belonging to the additional $  SU(2)'$ gauge
group which exists only in the Sh-world and has a large
confinement radius.

\end{frame}

\begin{frame}
\frametitle{The breaking $  E_6$ in the O- and Sh-
worlds}
\noindent
The breaking mechanism of the $  E_6$ unification is
given in Ref:

\vskip 0.2in
\noindent
The vacuum expectation values (VEVs) of the Higgs fields $
 H_{27}$ and $  H_{351}$ belonging to 27- and
351-plets of the $  E_6$ group can appear in the case
(iii) for the O-world only with nonzero 27-component:

$$    \langle H_{351}\rangle=0, \qquad v=\langle
H_{27}\rangle\neq 0.  $$
\noindent
In the case (iii) for the Sh-world we have

$$    \langle H_{27}\rangle=0, \qquad V=\langle
H_{351}\rangle\neq 0. $$

\end{frame}

\begin{frame}
\frametitle{The breaking $  E_6$ in the O- and Sh-
worlds}
\noindent
The 27 representation of $  E_6$ is decomposed into 1 +
16 + 10 under the $  SO(10)$ subgroup and the 27 Higgs
field $  H_{27}$ is expressed in 'vector' notation as
\bea
   H_{27}\equiv
    &\left(\begin{array}{c}H_0\\H_{\alpha}\\
                                          H_M\end{array}\right),
\eea
\noindent
where the subscripts 0, $  \alpha=1,2,...,16$ and $
 M=1,2,...,10$ stand for  singlet, the 16- and the
10-representations of $  SO(10)$, respectively. Then
\noindent
\bea  
   \langle H_{27}\rangle=
    &\left(\begin{array}{c}v\\0\\
                                          0\end{array}\right).
\eea
\noindent
Taking into account that the $   351-plet$ of $ 
E_6$ is constructed from $  27\times 27$ symmetrically,
we see that the trace part of $  H_{351}$ is a singlet
under the maximal little groups. Therefore, in a suitable basis,
we can construct the VEV $  \langle H_{351}\rangle$  for
the case of the maximal little group $  SU(2)\times
SU(6)$.

\end{frame}

\begin{frame}
\frametitle{The breaking $  E_6$ in the O- and Sh-
worlds}
\noindent
A singlet under this group which we get from a symmetric product
of $  27\times 27$ comes from the component $$ 
(1, 15)\times (1, 15)$$ and hence
\bea    \langle H_{351}\rangle
=&\left(\begin{array}{cc}V\otimes 1_{15}&\\&0\otimes
1_{15}\end{array}\right). \eea
\noindent
According to the assumptions of DLNT-paper, in the ordinary world
there exists the following chain of symmetry groups from the GUT
scale ($  M_{E6}$) of the $  E_6$ unification
down to the Standard Model (SM) scale ($  M_{EW}$):
$$  
E_6\to SO(10)\times U(1)_Z \to SU(4)_C\times SU(2)_L \times
SU(2)_R\times U(1)_Z$$ $$   \to SU(3)_C\times SU(2)_L
\times SU(2)_R\times U(1)_X\times
 U(1)_Z $$  
$$   \to [SU(3)_C\times SU(2)_L\times U(1)_Y]_{{SUSY}}\to
 SU(3)_C\times SU(2)_L\times U(1)_Y.
$$
\end{frame}

\begin{frame}
\frametitle{The breaking $  E_6$ in the O- and Sh-
worlds}
\noindent
In the shadow Sh-world, we have the following chain:
$$    E'_6 \to SU(6)'\times SU(2)'_{\theta}
\to SU(4)'_C\times SU(2)'_L\times SU(2)'_{\theta}\times U(1)'_Z $$
$$   \to SU(3)'_C\times
SU(2)'_L\times SU(2)'_{\theta}\times U(1)'_X \times U(1)'_Z$$
$$
  \to [SU(3)'_C\times SU(2)'_L\times
SU(2)'_{\theta}\times U(1)'_Y]_{{SUSY}}$$ $$\to SU(3)'_C\times
SU(2)'_L\times SU(2)'_{\theta}\times U(1)'_Y .$$
\noindent
In general, this is not an unambiguous choice of the $  
E_6(E'_6)$ breaking chains.
\noindent
The unification $  E_6=E'_6$ restores the mirror parity
MP at the scale:
\noindent
$$  M'_{{E_6}} = M_{{E_6}}\simeq 10^{18}\,\,{\mbox{GeV}}.$$

\end{frame}

\section{New shadow gauge group {$  SU(2)'_{\theta}$}}

\begin{frame}
\frametitle{New shadow gauge group $  SU(2)'_{\theta}$}
\noindent
Now we are confronted with the question:
\vskip 0.2in
\noindent
\setbeamercolor{mycolor2}{fg=blue,bg=green}
\begin{beamercolorbox}{mycolor2}\noindent
What group of symmetry $  SU(2)'$, unknown in the
O-world, exists in the Sh-world, ensuring the $  E'_6$
unification?
\end{beamercolorbox}
\vskip 0.2in

\begin{alertblock}{}
We assume that this new $  SU(2)'$ group is precisely the
$  SU(2)'_{\theta}$ gauge group of symmetry suggested by
\alert{L.B.~Okun.}

{  L. B.~Okun, JETP Lett. {\bf 31} (1980) 144; Pisma Zh.
Eksp. Teor. Fiz. {\bf 31} (1979) 156; Nucl. Phys. B {\bf 173}
(1980) 1.}
\end{alertblock}
\vskip 0.2in
\noindent
The reason for our choice of the $  SU(2)'_{\theta}$
group was to obtain the evolution $ 
{\alpha'}_{2\theta}^{-1}(\mu)$, which leads to the new scale in
the shadow world at extremely low energies.

\end{frame}

\begin{frame}
\frametitle{New shadow gauge group $  SU(2)'_{\theta}$}
\noindent
In the works by L.B.Okun it was suggested the hypothesis that in
Nature there exists the symmetry group
\noindent
$$   G_{\theta} = SU(3)_C\times SU(2)_L\times
SU(2)_{\theta}\times U(1)_Y \,,$$
\noindent
i.e. with an additional non-Abelian $   SU(2)_{\theta}$
group whose gauge fields are neutral, massless vector particles --
`thetons'.

\begin{alertblock}{}
These `thetons' have a macroscopic confinement radius $ 
1/\Lambda_{\theta}$.
\end{alertblock}
\end{frame}

\begin{frame}
\frametitle{New shadow gauge group $  SU(2)'_{\theta}$}
\noindent
Later, in Refs.:

\noindent
it was assumed that if there exists any $SU(2)$ group with the
scale
$$  \Lambda_2 \sim 10^{-3} \rm{eV},$$ then it is
possible to explain the small value of the observable $CC$.

\end{frame}

\begin{frame}
\frametitle{New shadow gauge group $  SU(2)'_{\theta}$}
\noindent
This idea was taken up in Refs.:

\noindent
Since this moment and further: Ref. DLT.
\noindent
Here we assume the existence of low-energy symmetry group $
 G_{\theta}$ ) only in the Sh-world:
\noindent
theta-particles are absent in the O-world, because their existence
is in disagreement with all experiments.
\noindent
However, they can exist in the Sh-world:
\noindent
$$    G' = SU(3)'_C\times SU(2)'_L\times
SU(2)'_{\theta}\times U(1)'_Y \,,$$
\noindent
Now we consider shadow thetons $  {\Theta'}^i_{\mu\nu}$,
$  i=1,2,3$, which belong to the adjoint representation
of the group $  SU(2)'_{\theta}$,
three generations of shadow theta-quarks $  q'_{\theta}$
and shadow theta- leptons $  l'_{\theta}$,
and the necessary theta-scalars $  \phi'_{\theta}$ for
the corresponding breakings.

\end{frame}

\begin{frame}
\frametitle{New shadow gauge group $  SU(2)'_{\theta}$}
\noindent
Shadow thetons have macroscopic confinement radius $ 
1/\Lambda'_{\theta}$, and we assume that
$$   \Lambda'_{\theta}\sim 10^{-3}\,\, {\mbox{eV}}.$$
\noindent
Matter fields  of the fundamental 27-representation of the $
 E'_6$ group decompose under $  SU(2)'_\theta\times
SU(6)'$ subgroup as follows: $$  27=(2,6)+(1,15),$$ 

\end{frame}

\begin{frame}
\frametitle{New shadow gauge group $  SU(2)'_{\theta}$}
\noindent
Where
\bea  
       (2,6) \to q' = &\left(\begin{array}{c}q'_{\theta,A}|_{I_{\theta}=+1/2}\\
                                          q'_{\theta,A}|_{I_{\theta}=-1/2} \end{array}\right).
                                          &\\
        (1,15) \to & D',D'^c&\\
                   &h' = \left(\begin{array}{c}h'^+\\
                                               h'^0
                                               \end{array}\right),
                                               &\\
                    &h'^{\rm c} = \left(\begin{array}{c}h'^0\\
                                               h'^-
                                               \end{array}\right),&\\
&{q'}^c_a, {N'}^c, S'.&
 \lb{4SW} \eea
Here $  A=1,...,6; a=1,2,3$ are color indices and $
 I_{\theta}$ is a theta-isospin; theta-quarks are $ 
q'_{\theta,A}$, while quarks $  {q'}^c_a$, right-handed
neutrino $  {N'}^c$ and scalar $S'$ are $ 
SU(2)'_{\theta}$-singlets.

\end{frame}

\section{The running of the inverse coupling constants}

\begin{frame}
\frametitle{The running of the inverse coupling constants}
\noindent
Let us consider now the running of the inverse coupling constants
in the SM and SM':
\vskip 0.2in
\noindent
$  \alpha_i^{-1}(\mu) = \frac{b_i}{2\pi}\ln
    \frac{\mu}{\Lambda_i},\quad \mbox{in the O-world},$

\noindent
$  {\alpha'}_i^{-1}(\mu) = \frac{b'_i}{2\pi}\ln
    \frac{\mu}{\Lambda'_i},\quad \mbox{in the M-world},$
\vskip 0.2in
\noindent
where $  \mu$ is the energy scale.
\vskip 0.2in
\noindent
And $  \alpha_i^{(\prime)} =\frac
{g^{(\prime)2}_i}{4\pi},    g^{(\prime)}_i$
is the gauge coupling constant of the gauge group $ 
G^{(\prime)}_i$.
\vskip 0.2in
\noindent
Here $  i=1,2,3$ correspond to $  U(1),\, SU(2)$
and $  SU(3)$ groups of the $  SM$( or $SM'$).

\noindent
Notations: here and below the ordinary world is given by the non-primed 
symbols, while mirror or shadow world is given by the primed symbols.
\end{frame}

\section{The running of coupling constants in the O- and Sh-worlds}

\begin{frame}
\frametitle{The running of coupling constants}
\noindent
In this section we consider the running of all the inverse gauge
coupling constants in the O- and Sh-worlds. 
\vskip 0.2in
\noindent
They are well described \underline{by the one-loop approximation}
of the renormalization group equations (RGEs), since from the
Electroweak (EW) scale up to the Planck scale ($
M_{Pl}$ ) all the non-Abelian gauge theories with rank
$  r\ge 2$ appearing in our model are chosen to be
{\color{red}asymptotically free}. 
\vskip 0.2in
\noindent
Considering Sh-world we have used the values of parameters $
\zeta$ and $ \xi$ estimated by
{\color{blue}Berezhiani-Dolgov-Mohapatra} from astrophysical
measurements:

\begin{block}{}
$$ \zeta= 30 \quad \rm{and} \quad
                         \xi= 1.5,\quad \xi= 
\frac{\Lambda^\prime_i}{\Lambda_i}$$
\end{block}
\end{frame}

\begin{frame}
\frametitle{The running of coupling constants}
\noindent
We start with the $  SM$ in the ordinary world:
$$  G_{{SM}} = SU(3)_C\times SU(2)_L\times U(1)_Y$$
and with $  SM'\times SU(2)'_\theta$ in the shadow world:
$$  G'_\theta= SU(3)'_C\times SU(2)'_L\times
SU(2)'_\theta\times U(1)'_Y. $$ Now we must consider the content of
particles for the $  SU(2)'_\theta$ gauge group.

\noindent
By analogy with the theory developed by L.B. Okun, we have shadow
thetons $ {\Theta'}^i_{\mu\nu}\,\,$ ($ 
i=1,2,3$), which belong to the adjoint representation of $
 SU(2)'_{\theta},$ three generations of shadow theta-quarks
$  q'_{\theta}$, shadow leptons $  l'_{\theta}$,
and theta-scalars $ \phi'_{\theta}$ as doublets of
$  SU(2)'_{\theta}$. 
\vskip 0.2in
\noindent
Calculating the slopes for the running $
{\alpha'_{2\theta}}^{-1}$: $  b_{2\theta} = 3 \quad
\rm{ and}\quad  b_{2\theta}^{SUSY}= -2, $ it is easy to obtain the
value:

\begin{alertblock}{}
$$  \alert{\Lambda'_{\theta}\sim 10^{-3}\,\,{\rm{eV}}}.$$
\end{alertblock}

\end{frame}

\section{$  MSSM^{(\prime)}$}

\begin{frame}
\frametitle{$  MSSM^{(\prime)}$}
\noindent
The Minimal Supersymmetric Standard Model ($ 
MSSM^{(\prime)}$) (which extends the conventional $ 
SM^{(\prime)}$) gives the running of the inverse coupling
constants from the supersymmetric scale $ 
M^{(\prime)}_{{SUSY}}$ up to the seesaw scale $ 
M^{(\prime)}_R$, where the heavy (shadow) right-handed neutrinos
are produced.
\vskip 0.2in
\noindent
In $  MSSM'$ the superpartners of particles, i.e., \alert{shadow
``sparticles''}, have the following heavy masses: $ \tilde
m' = \zeta \tilde m,$ and the supersymmetry breaking scale in the
Sh-world is larger:

\begin{block}{}
$$  M'_{{SUSY}} = \zeta M_{{SUSY}},$$
\end{block}
\vskip 0.2in
\noindent
but seesaw scale is the same as in the O-world:

\begin{block}{}
$$  M'_R = M_R.$$
\end{block}

\end{frame}

\begin{frame}
\frametitle{The running of coupling constants in the O-world}
\vskip 0.2in
\noindent
The running of the inverse coupling constants as functions of
$  x=\log_{10}\mu,$ where $  \mu$ is the energy
variable, is presented for O-world in {\bf Fig.~1,2}, using the
scales $  M_{{SUSY}}= 10\,\,\rm{TeV}$  and $ 
M_R =2.5\cdot 10^{14}\,\,\rm {GeV}.$ In these figures,
\underline{{\color{red}red} lines correspond to the ordinary
world}.
\vskip 0.2in
\noindent Here:
$$M_{GUT} =1.1\cdot 10^{16}\,\,\rm {GeV},$$
$$M^\prime_{GUT} =6.37\cdot 10^{17}\,\,\rm {GeV},$$
$$M_{E_6}\approx 6.96\cdot 10^{17}\,\,\rm {GeV},$$
$$ \alpha_{E_6}^{-1}\approx 27.64.$$
\vskip 0.2in
\noindent
{\bf Fig.~2} shows the running of the gauge coupling constants
near the scale of the $  E_6$  unification for $
 x\ge 15$.
\end{frame}

\begin{frame}
\frametitle{The running of coupling constants in the O-world}

\begin{figure} \centering
\includegraphics[height=50mm,keepaspectratio=true,angle=0]{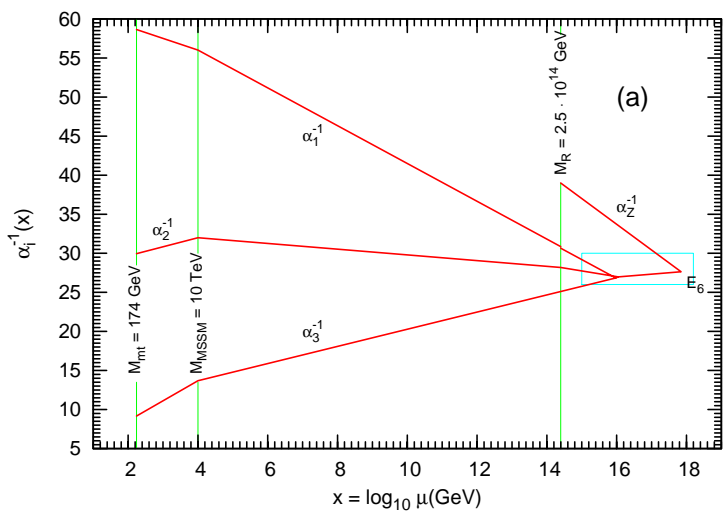}
\caption {This {\bf Fig.~1} presents the running of the inverse
coupling constants $ \alpha_i^{-1}(x)$ in the ordinary
world from the Standard Model up to the $  E_6$
unification for SUSY breaking scale $  M_{SUSY}= 10$ TeV
and seesaw scale $  M_R=2.5\cdot 10^{14}$ GeV. This case
gives: $  M_{E_6}\approx 6.96\cdot 10^{17}$ GeV and
$ \alpha_{E_6}^{-1}\approx 27.64$. }\end{figure}

\end{frame}

\begin{frame}
\frametitle{The running of coupling constants in the O-world}

\begin{figure} \centering
\includegraphics[height=50mm,keepaspectratio=true,angle=0]{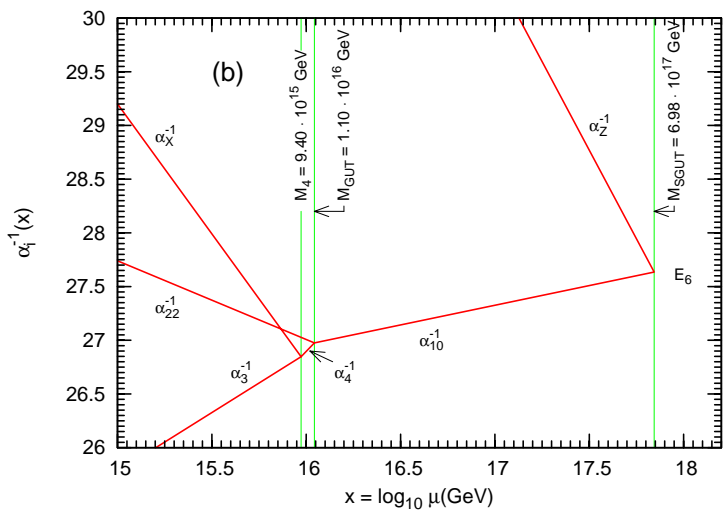}
\caption {This {\bf Fig.~2} is the same as {\bf Fig.~1}, but
zoomed in the scale region from $  10^{15}$ GeV up to the
$  E_6$ unification to show the details.}\end{figure}

\end{frame}

\begin{frame}
\frametitle{The running of the coupling constants in the Sh-world}
\noindent
The running of the inverse coupling constants as functions of
$  x=\log_{10}\mu$ is presented for Sh-world in {
Fig.~3,4}, using the scales $  M'_{{SUSY}}= \zeta
M_{SUSY} = 300\,\,\rm{TeV}$ and $  M'_R = M_R = 2.5\cdot
10^{14}\,\, \rm {GeV}.$ In these figures,
\underline{{\color{blue}blue} lines correspond to the shadow
world}.
\vskip 0.2in
\noindent
{\bf Fig.~4} shows the running of the gauge coupling constants
near the scale of the $  E_6$ unification for $ 
x\ge 15$.

\begin{alertblock}{}
Here we see the running of the $ \theta$-coupling
constant $ {\alpha'_{2\theta}}^{-1}$.
\end{alertblock}

\end{frame}

\begin{frame}
\frametitle{The running of the coupling constants in the Sh-world}

\begin{figure} \centering
\includegraphics[height=50mm,keepaspectratio=true,angle=0]{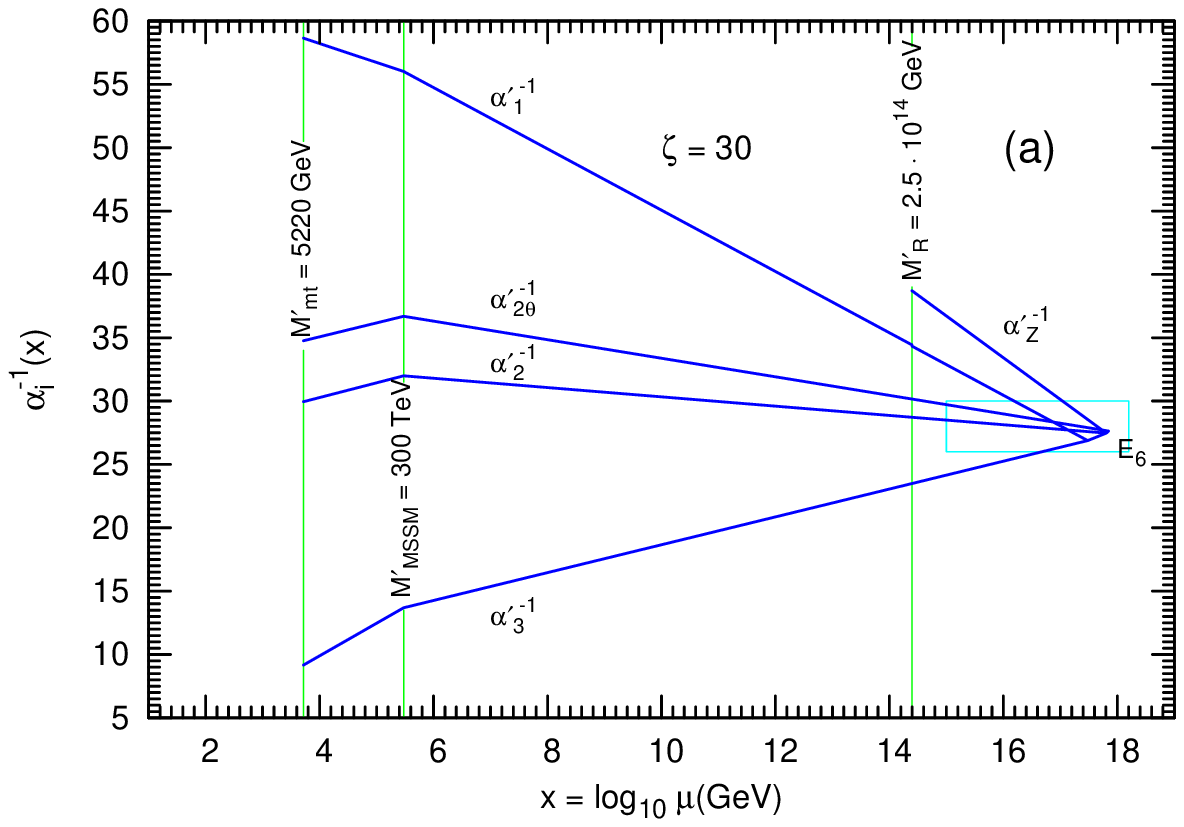}
\caption {This {\bf Fig.~3} presents the running of the inverse
coupling constants $ \alpha_i^{-1}(x)$ in the shadow
world from the Standard Model up to the $  E_6$
unification for shadow SUSY breaking scale $  M'_{SUSY}=
300$ TeV and shadow seesaw scale $  M'_R=M_R= 2.5\cdot
10^{14}$ GeV; $ \zeta = 30$. This case gives: $ 
M_{E_6}= 6.96\cdot 10^{17}$ GeV and $
\alpha_{E_6}^{-1}=27.64$.}
\end{figure}

\end{frame}

\begin{frame}
\frametitle{The running of the coupling constants in the Sh-world}

\begin{figure} \centering
\includegraphics[height=50mm,keepaspectratio=true,angle=0]{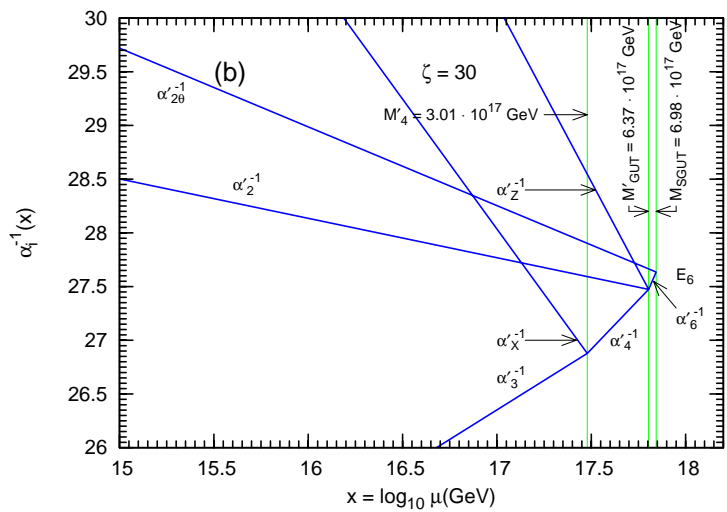}
\caption {This {\bf Fig.~4} is the same as {\bf Fig.~3}, but
zoomed in the scale region from $  10^{15}$ GeV up to the
$  E_6$ unification to show the details.}
\end{figure}

\end{frame}

\begin{frame}
\frametitle{The running of the coupling constants in the Sh-world}
\noindent
The comparison of the evolutions in the O- and Sh-worlds is
presented in {\bf  Figs.~5,6.}
\vskip 0.2in
\noindent
The parameters of our model are as follows: $ 
M_{SUSY}=10\,\,{\mbox{TeV}},$
 $ \zeta =30.$
In this case we have: $  M'_{SUSY}=300 \,\,{\mbox
{TeV}},$ $  M_R=M'_R=2.5\cdot 10^{14}\,\,{\mbox {GeV}},$
Here $  M_{GUT}=1.10\cdot 10^{16}\,\,{\mbox
{GeV}}\qquad\to {\mbox {for}}\,\,SO(10),$ $ 
M'_{GUT}=6.37\cdot 10^{17}\,\,{\mbox {GeV}}\qquad\to {\mbox
{for}}\,\,SU'(6),$ $  M_{E_6}= 6.96\cdot
10^{17}\,\,{\mbox {GeV}},$ and $ \alpha_{E_6}^{-1}=27.64.
$

\end{frame}

\begin{frame}
\frametitle{The running of the coupling constants in the Sh-world}

\begin{figure} \centering
\includegraphics[height=50mm,keepaspectratio=true,angle=0]{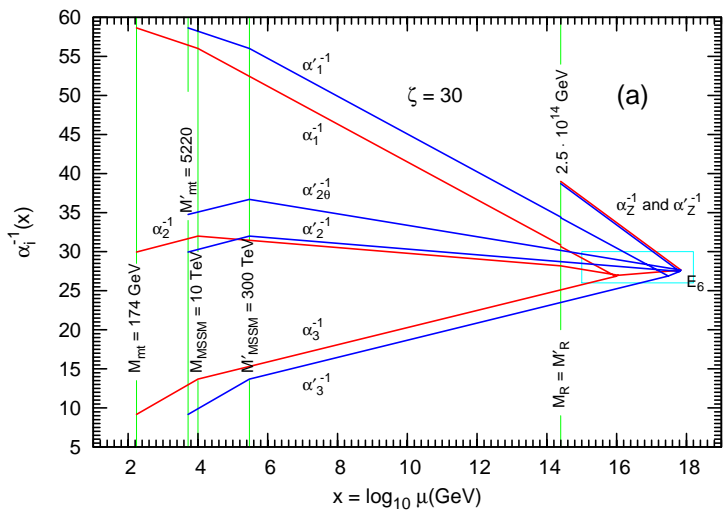}
\caption {This {\bf Fig.~5} shows the running of the inverse
coupling constants $ \alpha_i^{-1}(x)$ in both ordinary
and mirror worlds with broken mirror parity, from the Standard
Model up to the $  E_6$ unification for SUSY breaking
scales $  M_{SUSY}= 10$ TeV, $  M'_{SUSY}= 300$
TeV and seesaw scales $  M_R=M_R'=2.5\cdot 10^{14}$ GeV,
$ \zeta = 30$. This case gives: $ 
M_{E_6}=6.96\cdot 10^{17}$ GeV and $
\alpha_{E_6}^{-1}=27.64$. }
\end{figure}

\end{frame}

\begin{frame}
\frametitle{The running of the coupling constants in the Sh-world}

\begin{figure} \centering
\includegraphics[height=50mm,keepaspectratio=true,angle=0]{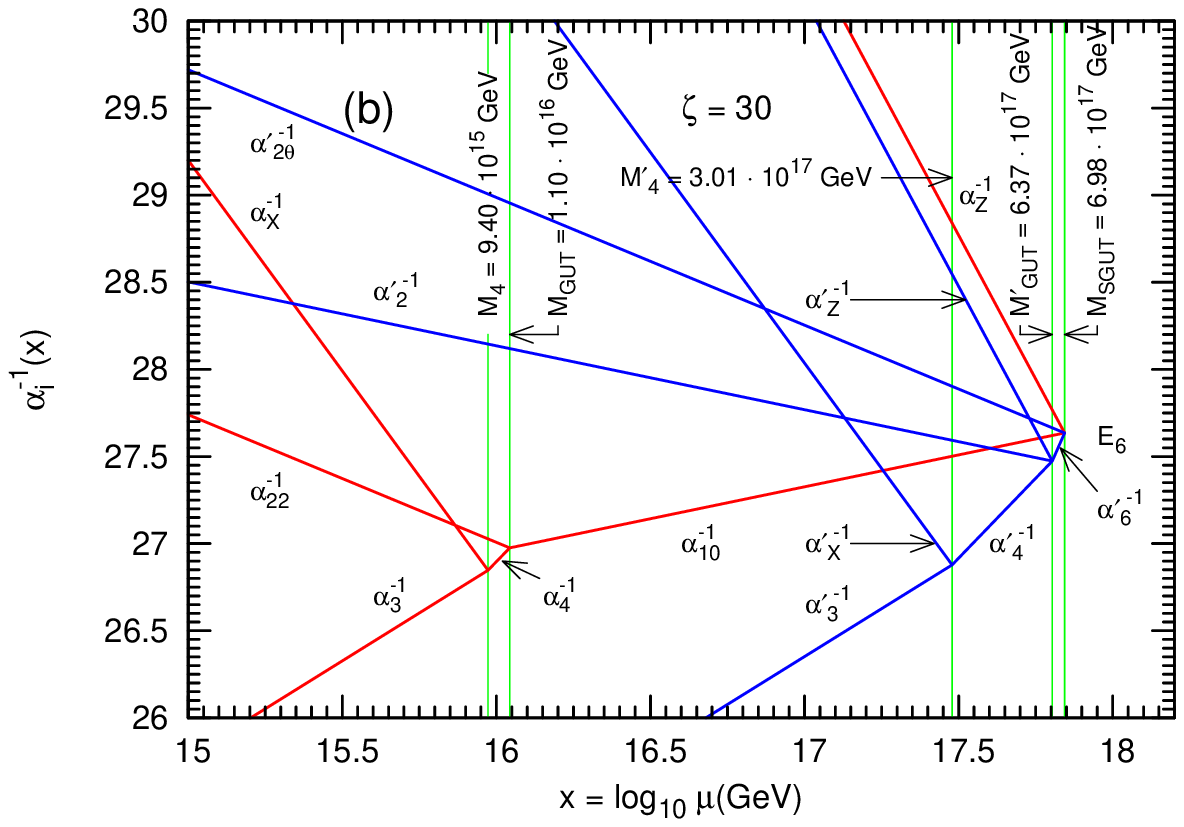}
\caption {This  {\bf Fig.~6} is the same as {\bf Fig.~5}, but
zoomed in the scale region from $  10^{15}$ GeV up to the
$  E_6$ unification to show the details.}
\end{figure}

\end{frame}

\section{Inflation, $ E_6$ unification and the problem of walls in the Universe}

\begin{frame}
\frametitle{Inflation, $ E_6$ unification and the problem of
walls}
\noindent
The simplest model of inflation is based on the superpotential:
$$    W=\lambda \varphi(\Phi^2-\mu^2).$$
\vskip 0.2in
\noindent
It contains the inflaton field given by $\varphi$ (singlet of
$  E_6$) and the Higgs field $  \Phi$.
\vskip 0.2in
\noindent
$  \lambda$ is a coupling constant of order 1
and $  \mu$ is a dimensional parameter of the order of
the GUT scale:
\vskip 0.2in
\noindent
See, for example,

\vskip 0.2in
\noindent
The supersymmetric vacuum is located at $  \varphi=0$,
$  \Phi=\mu$.
\end{frame}

\begin{frame}
\frametitle{Inflation, $ E_6$ unification and the problem of
walls}
\noindent
But for the field values $  \Phi=0$, $ 
|\varphi| > \mu$ the tree level potential has a flat valley with
the energy density $  V=\lambda^2\mu^4$.
\vskip 0.2in
\noindent
When the supersymmetry is broken by the non-vanishing F-term, the
flat direction is lifted by radiative corrections and the inflaton
potential acquires a slope appropriate for the slow roll
conditions.
\vskip 0.2in
\noindent
This so-called hybrid inflation model leads to the choice of the
initial conditions:

\end{frame}

\begin{frame}
\frametitle{Inflation, $ E_6$ unification and the problem of
walls}
\noindent
Namely, at the end of the Planck epoch the singlet scalar field
$\varphi$ should have an initial value
$$  \varphi=f\sim 10^{18}\,\,{\rm{GeV}}$$
e.g. $  \sim E_6$-GUT scale,
while the field $  \Phi$ must be zero with high accuracy
over a region much larger than the initial horizon size $
 \sim M_{Pl}$.
\vskip 0.2in
\noindent
In other words, the initial field configuration should be located
right on the bottom of the inflaton valley and the energy density
starts with $$  V=\lambda^2\mu^4 \ll M_{Pl}^4.$$
\noindent
If $  E_6'$ is the mirror counterpart of $ 
E_6$, then we have $  Z_2$ symmetry, i.e. a discrete
group connected with the mirror parity.

\begin{block}{}  In general, the spontaneous breaking of a discrete
group leads to phenomenologically unacceptable walls of huge
energy per area.
\end{block}
\noindent
{\bf  Fig.~7} demonstrates this situation.

\end{frame}

\begin{frame}
\frametitle{Inflation, $ E_6$ unification and the problem of
walls}

\begin{center}
\scalebox{1} 
{
\begin{pspicture}(0,-4.004841)(11.174835,3.9848409)
\psbezier[linewidth=0.04,shadow=true,shadowangle=-45.0,fillstyle=gradient,gradlines=2000,gradbegin=red,
gradend=blue,gradmidpoint=1.0,gradangle=44.999996](1.1399997,2.7989092)(0.67250633,1.9149126)
(3.2434227,-1.1982028)(4.14,-1.6410909)(5.0365767,-2.083979)(10.0,-2.1810908)(10.24,-1.2810909)
(10.48,-0.38109082)(7.4934225,2.4118967)(6.58,2.8189092)(5.6665773,3.2259216)(1.607493,3.6829057)
(1.1399997,2.7989092)
\psbezier[linewidth=0.04,shadow=true,shadowangle=-45.0,fillstyle=gradient,gradlines=2000,
gradbegin=magenta,gradmidpoint=1.0,gradangle=90.0](0.58131564,-2.1189015)(1.1626313,-2.9810908)
(5.4029098,-1.4745029)(6.28339,-0.9561969)(7.1638703,-0.43789086)(11.08,2.1189091)
(9.852628,3.041875)(8.625256,3.964841)(5.2210903,2.7774158)(4.2657266,2.2189093)
(3.3103628,1.6604025)(0.0,-1.2567121)(0.58131564,-2.1189015)
\psbezier[linewidth=0.08,linecolor=green,doubleline=true,doublesep=0.12,doublecolor=yellow]
(5.58,3.7989092)(5.66,2.918909)(6.229719,3.9476376)(6.1,2.9789093)(5.97028,2.0101807)
(5.446328,2.2041657)(5.12,1.2589092)(4.793671,0.31365272)(3.9852622,-2.6786351)(4.9799995,-2.7810907)
\rput(2.897656,2.7339091){ \color{white}$H_{27}$ on O-brane}
\rput(7.957656,-1.3460908){ \color{white}$H_{27}$ on
Sh-brane}
\rput(2.5176558,-1.4660908){ $H_{351}$ on O-brane}
\rput(7.997656,2.4939091){ $H_{351}$ on Sh-brane}
\rput(6.8185935,1.0389092){ Wall}
\psline[linewidth=0.06cm,arrowsize=0.05291667cm
4.0,arrowlength=2.0,arrowinset=0.4]{->}(6.14,0.9389092)(5.2,0.39890918)
\rput(5.106875,-3.7210908){ Fig. 7}
\end{pspicture}
}
\end{center}

\end{frame}

\begin{frame}
\frametitle{Inflation, $ E_6$ unification and the problem of
walls}
\noindent
Then we have the following properties for the energy densities of
radiation, DM, M and wall:
$$   \rho_r\varpropto \frac 1{a(t)^4}, \quad \rho_{M,DM}\varpropto
\frac 1{a(t)^3}, \quad \rho_{wall}\varpropto \frac 1{a(t)}, $$
where $  a(t)$ is a scale factor with cosmic time $
 t$ in the Friedmann-Lemaitre-Robertson-Walker (FLRW) metric
describing our Universe.
\vskip 0.2in
\noindent
For large Universe we have $ 
\rho_{wall}\gg\rho_{M,DM},\rho_r.$
\vskip 0.2in
\noindent
In our case of the hidden world, the shadow superpotential is:
$$   W'=\lambda' \varphi'({\Phi'}^2-\mu'^2), $$
where $  \Phi'=H_{351}$ and $  \langle
H_{351}\rangle=\mu'$.

\end{frame}

\begin{frame}
\frametitle{Inflation, $ E_6$ unification and the problem of
walls}
\noindent
Then the initial energy density in the Sh-world is
$$ 
V'={\lambda'}^2{\mu'}^4 \ll M_{Pl}^4.$$
To avoid this phenomenologically unacceptable wall dominance we
cannot assume symmetry under $  Z_2$ and thus $ 
V=V'$ is not automatic.
\vskip 0.2in
\noindent
Instead, it is necessary to assume the following finetuning:
$$    V=V': \quad \lambda^2\mu^4={\lambda'}^2{\mu'}^4,$$
which helps to obtain the initial conditions for the GUT-scales
and GUT-coupling constants: $$  M_{E6}=M'_{E6'},$$
$$  g_{E6}=g'_{E6'}.$$

\end{frame}

\section{The cosmological constant problem}

\begin{frame}
\frametitle{The cosmological constant problem }
\noindent
The cosmological constant ($CC$) was first introduced by Einstein in
1917  with aim to admit a static cosmological solution in his new
general theory of relativity.
\vskip 0.2in
\noindent
The the bare cosmological constant, $\lambda$ was accomplished by 
the addition to the original field equations:
$$    G^{\mu\nu}= R^{\mu\nu} - \frac{1}{2}Rg^{\mu\nu}=
8\pi GT^{\mu\nu}    $$
of the divergence-free term $   -\lambda g_{\mu\nu}$:
$$   G^{\mu\nu}= 8\pi GT^{\mu\nu} -\lambda g^{\mu\nu},
 $$
where
$  R_{\mu\nu}$ is the Ricci curvature of $ 
g_{\mu\nu}$,
and $  T_{\mu\nu}$ is the energy-momentum tensor of
matter.

\end{frame}

\begin{frame}
\frametitle{The cosmological constant problem }
\noindent
Later it was realized:
\vskip 0.2in
\noindent
 See

\noindent
that quantum fluctuations result in a vacuum energy, $ 
\rho_{vac}$:
\vskip 0.2in
\noindent
any mode contributes $  \frac{1}{2}\hslash \omega$ to the
vacuum energy, and the expected value of the energy momentum
tensor of matter is:
$$   \langle T^{\mu\nu}\rangle = T^{\mu\nu}_m -
\rho_{vac}g^{\mu\nu}, $$
where $  T^{\mu\nu}_m$ vanishes in vacuum.

\end{frame}

\begin{frame}
\frametitle{The cosmological constant problem }
\noindent
The quantum expectation of the energy-momentum tensor, $ 
\langle T^{\mu\nu}\rangle$, acts as a source for the Einstein
tensor:
$$  G^{\mu\nu}= 8\pi GT^{\mu\nu}_m -\Lambda g^{\mu\nu},
 $$
where $  \Lambda$ is the effective cosmological
constant provided by the contribution of the vacuum energy,
$  \rho_{vac}$.
\vskip 0.2in
\noindent
We would expect that the effective vacuum energy:
$$   \rho_{vac}^{(eff)} = \frac{\lambda}{8\pi G} +
\rho_{vac} = \frac{\Lambda}{8\pi G}  $$
to be no smaller than $  \rho_{vac}$.
\vskip 0.2in
\noindent
Even if the ``bare'' cosmological constant is assumed to vanish
($\lambda=0$), the effective cosmological constant is not equal to
zero.

\end{frame}

\begin{frame}
\frametitle{The cosmological constant problem }
\noindent
Requirement that $  \Lambda = 0$ means that there must be
an exact cancellation between the 'bare' cosmological constant,
$  \lambda$, and the vacuum energy stress, $ 
8\pi G\rho_{vac}$:
$$   \Lambda=0 \quad \to \quad  \lambda + 8\pi
G\rho_{vac}=0.  $$
When the spontaneous symmetry breaking was widely discussed in the
Standard Model, Veltman commented that the vacuum energy arising
in spontaneous symmetry breaking gives an additional contribution
to the $CC$:

\noindent
If we assume that the field theory is only valid up to some energy
scale $  M_{cutoff}$, then there is a contribution to
$  \rho_{vac}$ of $O(M_{cutoff}^4)$.\\
\vskip 0.2in
\noindent
Collider experiments have established that the SM is accurate up
to energy scales $  M_{cutoff}\gtrsim O(M_{EW})$, where
$  M_{EW}\approx 246$ GeV is the EW-scale. We would
therefore expect $  \rho_{vac}$ to be at least $
 O(M_{EW}^4)$.
\end{frame}

\begin{frame}
\frametitle{The cosmological constant problem }
\noindent
In the absence of any new physics between the electroweak and the
Planck scale, $  M_{Pl} \approx 1.2\times 10^{19}$ GeV,
where quantum fluctuations in the gravitational field can no
longer be safely neglected, we would expect $ 
\rho_{vac}\sim O(M^4_{Pl})$. If supersymmetry were an unbroken
symmetry of Nature, the quantum contributions to the vacuum energy
would all exactly cancel leaving $  \rho_{vac}=0$ and
$  \Lambda=\lambda$. However, our universe is not
supersymmetric today, and so SUSY must have been broken at some
energy scale $  M_{SUSY}$, where $  1\,\,
{\rm{TeV}}\lesssim M_{\rm{SUSY}}\lesssim M_{Pl}$. It is necessary
to comment that the SUSY breaking is necessary in our superstring
and thereby SUSY-based model. We would expect $ 
\rho_{vac}\sim O(M^4_{\rm{SUSY}})$. Our model of quantum cosmology
also had to take into account extra dimensions and branes,
spontaneous breaking of compactification.

\end{frame}

\begin{frame}
\frametitle{The cosmological constant problem }
\noindent
Previously in Refs.

\end{frame}

\begin{frame}
\frametitle{The cosmological constant problem }
\noindent
It was shown that SUGRA models which ensure the vanishing of the
vacuum energy density near the physical vacuum lead to a natural
realization of {\bf   the Multiple Point Model (MPP)}:

\noindent
MPP describes the degenerate vacua with $\Lambda=0$.
\end{frame}

\begin{frame}
\frametitle{The cosmological constant problem }
\noindent
The expansion rate of our Universe is sensitive to $ 
\rho_{vac}^{(eff)}$, or equivalently $  \Lambda$.
\vskip 0.2in
\noindent
The result of astrophysical measurements is given by:

$$  {(\rho_{vac}^{(eff)}})^{1/4}\simeq 2.3\times 10^{-3}
{\rm{eV}}$$ This implies that $\rho_{vac}^{(eff)}$ is some
$10^{60} - 10^{120}$ times smaller than the expected contribution
from quantum fluctuations, and gives rise to the cosmological
constant problem:

\begin{alertblock}{} ``Why is the measured effective
vacuum energy or cosmological constant so much smaller than the
expected contributions to it from quantum fluctuations?"
\end{alertblock}

\end{frame}

\section{A proposal for solving the $CC$ problem}

\begin{frame}
\frametitle{A proposal for solving the $CC$ problem}
\noindent
Here we follow the ideas  which gives a possible way to solve the
$CC$ problem:

\noindent
In quantum mechanics we consider the probability amplitudes: The
initial state $  |I\rangle$ transforming to a final state
$  |F\rangle$. In this spirit, using the Euclidian action
$  S_E$, only with the Ricci scalar $  R$ and $CC$
$  \Lambda$,
E.~Baum and S.~Hawking  have calculated the path integral in the
Euclidian space-time:

\noindent
which gives the following expression:
$$
        e^{-S_E} = e^{3\pi M_{Pl}/\Lambda}.   $$
So, $  \Lambda = 0$ dominates the action integral, which
is interpreted as the probability for $\Lambda = 0$ is close to 1.

\end{frame}

\begin{frame}
\frametitle{A proposal for solving the $CC$ problem}
\noindent
The essence of the new approach is that the bare cosmological
constant $    \lambda$ is promoted from a parameter to a
field.
\vskip 0.2in
\noindent
The minimization of the action with respect to $  
\lambda$ then yields an additional field equation, which
determines the value of the effective $CC$, $    \Lambda$.
In the classical history it dominates the partition function of
the Universe, $    Z$.
\vskip 0.2in
\noindent
If we take the total action of the Universe:
$$   S_{tot}(g_{\mu\nu},
\Psi^a, \Lambda;{\cal M}),$$
defined on a manifold $    \cal M$,
and with effective cosmological constant $\Lambda$,
$   \Psi^a$ are the matter fields
and $  g_{\mu\nu}$ is the metric field.

\end{frame}

\begin{frame}
\frametitle{A proposal for solving the $CC$ problem}
\noindent
Then we define $  S_{class}(\Lambda;{\cal M})$ to be the
value of $  S_{tot}(g_{\mu\nu},\Psi^a,\Lambda;{\cal M})$
evaluated with $   g_{\mu\nu}$  and  $   \Psi^a$
obeying their classical field equations for fixed boundary initial
conditions, and obtain the field equation for the effective $CC$
$  = \Lambda$, given by
$$      \frac{dS_{class}(\Lambda;{\cal M})}{d\Lambda} =
0.  $$
If $  \Lambda \approx 0$ dominates the action integral,
then we have an approximate cancellation between the 'bare'
cosmological constant and the vacuum energy stress:
$$   \Lambda \approx 0 \quad \to \quad  \lambda \approx
- 8\pi G\rho_{vac}.  $$
\end{frame}

\section{Dark energy: Quintessence model of cosmology}

\begin{frame}
\frametitle{Dark energy: Quintessence model of cosmology}
\noindent
Quintessence is described by a complex scalar field $ 
\varphi$ minimally coupled to gravity.
\vskip 0.2in
\noindent
 In our theory $ 
\varphi$ is a singlet of $  E_6$.
\vskip 0.2in
\noindent
The dynamics of two worlds, ordinary and hidden, is governed by
the following action:
$$    S= \int d^4x \sqrt{-g} \left[\frac
{1}{2\kappa^2}R + \lambda + {(\nabla \varphi)}^2 - V(\varphi) + L
+ L' + L_{mix}\right],  $$
where
$$    {(\nabla \varphi)}^2 =
g^{\mu\nu}\partial_{\mu}\varphi
\partial_{\nu}\varphi,   $$
and $   V(\varphi)$ is the potential of the field $
  \varphi$,

\noindent
$   \kappa^2=8\pi G=M_{Pl}^{-2}$,
$   M_{Pl}$ is the reduced Planck mass,

\noindent
$   R$ is the space-time curvature,
$   \lambda$ is a `bare' cosmological constant,

\noindent
$   L(L')$ is the Lagrangian of the O-(Sh-) sector,

\end{frame}

\begin{frame}
\frametitle{Dark energy: Quintessence model of cosmology}
\noindent
and $  L_{mix}$ is the Lagrangian of photon-photon$'$,
neutrino-neutrino$'$, etc. mixing:

\vskip 0.2in
\noindent
When both $  E_6$ and $  E'_6$ symmetry groups
are broken, then  down to $   G_{SM}$ and $  
G'_{SM}\times SU(2)'_\theta$ subgroups, respectively, we have:
$$   L = L_{gauge} +  L_{Higgs} +  L_{Yuk},$$
$$   L' =  L'_{\theta} + L'_{gauge} +  L'_{Higgs} +  L'_{Yuk},$$
$$   L_{tot} = L + L'+ L_{mix}.$$

\end{frame}

\begin{frame}
\frametitle{Dark energy: Quintessence model of cosmology}
\noindent
The two sectors mean that at least below the scales $ 
M_R=M'_R$ the degrees of freedoms (fields) can be classified into
fields from section O and fields from section H (hidden).

\noindent
Since this moment and further: $ H(hidden)\equiv Sh(shadow)$.
\vskip 0.2in
\noindent
 We could thus
consider the energy density due to zero point fluctuations in
\\the H-fields as contributing to $  \rho^{(H)}_{vac}$\\
while the O-fields contribute to $  \rho^{(O)}_{vac}.$ 
\vskip 0.2in
\noindent
Here we see that
$$  \rho^{(O)}_{vac}= \rho^{(SM)}_{vac},
  $$
and
$$  \rho^{(H)}_{vac}= \rho^{(SM')}_{vac} +
\rho^{(\theta)}_{vac}.
  $$
$$     \rho_{vac} =  \rho^{(O)}_{vac} + \rho^{(H)}_{vac} + \rho^{(mix)}_{vac}. $$
We can neglect $   \rho^{(mix)}_{vac}$.

\end{frame}

\begin{frame}
\frametitle{Dark energy: Quintessence model of cosmology}
\noindent
Then assuming that the 'bare' cosmological constant $ 
\lambda$ compensates the contribution of the SM and SM',e.g.
$$   \rho^{(SM-SM')}_{vac} = \rho^{(SM)}_{vac} + \rho^{(SM')}_{vac}$$
giving
$$    \lambda + 8\pi G\rho^{(SM-SM')}_{vac}=0.  $$
Then the effective $CC$, $   \Lambda$, is not zero:
 $$   \Lambda = 8\pi
G\rho^{(\theta)}_{vac},  $$
and the effective vacuum energy density is equal to DE density:
$$     \rho_{DE} = \rho^{(eff)}_{vac} =
\rho^{(\theta)}_{vac}.  $$
This speculative consideration explains a tiny value of the DE
density calculated into the next Section.

\end{frame}

\section{Dark Energy. Inflaton, axion and DE density}

\begin{frame}
\frametitle{Dark energy: Inflaton, axion and DE density}
\noindent
We assume that there exists an axial $  U(1)_A$ global
symmetry in our theory, which is spontaneously broken at the scale
$  f$ by a singlet complex scalar field $ 
\varphi$:
$$   \varphi = (f + \sigma) \exp(ia_{ax}/f). $$
We assume that a VEV $$  \langle \varphi \rangle = f$$ is
of the order of the $  E_6$ unification scale: $
 f\sim 10^{18}$ GeV.
\vskip 0.2in
\noindent
The real part $  \sigma $ of the field $ 
\varphi$ is the $\un {   inflaton}$,
while the boson $  a_{ax}$ (imaginary part of the singlet
scalar fields $  \varphi$) is an $\un {  axion}$
and could be identified with the massless Nambu-Goldstone (NG)
boson if the corresponding $  U(1)_A$ symmetry is not
explicitly broken by the gauge anomaly.

\end{frame}

\begin{frame}
\frametitle{Dark energy: Inflaton, axion and DE density}
\noindent
However, in the hidden world the explicit breaking of the global
$  U(1)_A$ by $  SU(2)'_\theta$ instantons
inverts $  a_{ax}$ into a pseudo Nambu-Goldstone (PNG)
boson $  a_{\theta}$. Therefore, in the Sh-world we have:
$$  \varphi' = (f + \sigma') \exp(ia_{\theta}/f).
 $$
\vskip 0.2in
\noindent
The flat FLRW spacetime gives the following field equation for
axion $  a_{\theta}$ :
$$   \frac{d^2 a_\theta}{dt^2} + 3 H \frac{d
a_\theta}{dt} + V'(a_\theta) = 0.  $$
where $  H$ is the Hubble parameter.
\end{frame}

\begin{frame}
\frametitle{Dark energy: Inflaton, axion and DE density}
\noindent
The singlet complex scalar field $  \varphi$ reproduces a
Peccei-Quinn (PQ) model.  Near the vacuum, a PNG mode
$  a_\theta$ emerges the following PQ axion potential:
$$  V_{PQ}(a_\theta) \approx {(\Lambda'_{\theta})}^4
         \left(1 - \cos(a_\theta/f)\right).   $$
\vskip 0.2in
\noindent
This axion potential exhibits minima at
$$  {(V_{PQ})}|_{min} = 0,  $$
where:
$$  \cos(a_\theta/f) = 1,\quad \rm{i.e.}\quad
{(a_\theta)}_{min}= 2\pi n f, \quad n = 0,1,... $$

\end{frame}

\begin{frame}
\frametitle{Dark energy: Inflaton, axion and DE density}
\noindent
For small fields $  a_\theta$ we expand the effective PQ
potential near the minimum:
$$   V_{PQ}(a_\theta) \approx
\frac{({\Lambda'_{\theta})}^4}{2f^2} (a_\theta)^2 + ... = \frac 12
m^2 {(a_\theta)}^2 + ..., $$
and hence the PNG axion mass squared is given by:
$$  m^2\sim {(\Lambda'_{\theta})}^4/f^2.  $$
Solving equation for  $  a_\theta$ we can use the axion
potential:
$$  V(a_\theta)=V_{PQ}(a_\theta),  $$
which gives:
$$  V'(a_\theta)
=\frac{{(\Lambda'_\theta)}^4}{f}\sin(a_{\theta}/f).  $$
If now $  \sin(a_\theta/f)=0$,

then $  {\dot a}_\theta =0$,

and $  V_{PQ}(a_\theta)=0$,

because $  \cos(a_\theta/f)=1$, according to minima.

\end{frame}

\begin{frame}
\frametitle{Dark energy: Inflaton, axion and DE density}
\noindent
The minimum of the total $  \theta$-potential is:
$$  V_{\theta}|_{min} = V_{PQ}(a_\theta)|_{min} +
V_{\theta-condensate}. $$
The first term is zero and we obtain:
$$   V_{\theta-condensate} = {(\Lambda'_\theta)}^4.
 $$
In this case when $  a_\theta={\rm{const}}$ and $
 {\dot a}_\theta =0 $, the contribution of axions to the energy
density of the Sh-sector is equal to zero.
\vskip 0.2in
\noindent
Finally, we obtain:
$$   \rho^{(eff)}_{vac} = \rho^{(\theta)}_{vac} =
|{\dot a}_\theta|^2 + V_{\theta}|_{min} = {(\Lambda'_\theta)}^4.
   $$

\end{frame}

\begin{frame}
\frametitle{Dark energy: Inflaton, axion and DE density}
\noindent
The DE density is equal to the value:
$$  \rho_{DE} =  \rho^{(eff)}_{vac} =
{(\Lambda'_\theta)}^4.  $$
Taking into account the results  of recent astrophysical
observations, we obtain the estimate of the $ 
SU(2)'_\theta$ group's gauge scale:
$$  \Lambda'_\theta \backsimeq  2.3\times
10^{-3}\,\,\rm{eV}.  $$
If $  \Lambda'_\theta \sim 10^{-3}$ eV and $ 
f\sim 10^{18}$ GeV,  we can estimate the $  \theta$-axion
mass:
$$  m\sim {\Lambda'_{\theta}}^2/f\sim 10^{-42}
\,\,{\rm{GeV}}. $$
It is extremely small.
\vskip 0.2in
\noindent
We have seen that these light axions do not give the contribution
to $  \rho_{DE}$. It is given only by the condensate of
$  \theta$-fields.

\end{frame}

\section{Reheating and radiation}

\begin{frame}
\frametitle{Reheating and radiation}
\noindent
The two sectors, ordinary and hidden, have different cosmological
evolutions. In particular, they never had to be in equilibrium
with each other: the Big Bang Nucleosynthesis (BBN) constraints
require that Sh-sector must have smaller temperature than
O-sector: $   T'<T$.
\vskip 0.2in
\noindent
See

\noindent
Since this moment and further: Ref. BKPRRS.
\vskip 0.2in
\noindent
During reheating the exponential expansion, which was developed by
inflation, ceases and the potential energy of the inflaton field
decays into a hot relativistic plasma of particles. At this point,
the Universe is dominated by radiation, and then quarks and
leptons are formed.
\end{frame}

\begin{frame}
\frametitle{Reheating and radiation}
\noindent
All the difference between the O- and Sh- worlds can be described
in terms of two macroscopic (free) parameters of the model:
$$   x\equiv \frac{T'}{T}, \quad \beta\equiv
\frac{\Omega'_B}{\Omega_B},  $$
where $   T(T')$ is O-(Sh-) photon temperature in the
present Universe, and $\Omega_B(\Omega'_B)$ is O-(Sh-)baryon
fraction.
\vskip 0.2in
\noindent
The modern observational data indicate that the Universe is almost
flat, in a perfect accordance with the inflationary paradigm.
\vskip 0.2in
\noindent
The relativistic fraction is represented by photons and neutrinos.

\end{frame}

\begin{frame}
\frametitle{Reheating and radiation}
\noindent
The contribution of the Sh-degrees of freedom to the observable
Hubble expansion rate, which are equivalent to an effective number
of extra neutrinos $$   \Delta N_{\nu}=6.14\cdot x^4,$$
is small enough:

$$   \Delta N_{\nu}=0.05\quad{\rm for}\quad x=0.3.$$ 
\vskip 0.2in
\noindent
In our model:
$$   \omega_r=\Omega_r h^2=4.2\cdot
10^{-5}(1+x^4),\quad h=\frac{H}{H_0},  $$
where the contribution of Sh-species is negligible due to the BBN
constraint: $   x^4\ll 1$.

\end{frame}

\begin{frame}
\frametitle{Reheating and radiation}
\noindent
Recent cosmological observations  show that for redshifts $$
  (1+z) \gg 1$$ we have:
$$      H(z) = H_0[\Omega_r(1+z)^4 + \Omega_m(1+z)^3].
$$
Therefore, the radiation is dominant at the early epochs of the
Universe, but it is negligible at present epoch: $$  
\Omega_r^{(0)}\ll 1.$$

\end{frame}

\begin{frame}
\frametitle{Reheating and radiation}
\noindent
Any inflationary model have to describe how the SM-particles were
generated at the end of inflation. The inflaton, which is a
singlet of $   E_6$, can decay, and the subsequent
thermalization of the decay products can generate the
SM-particles. The inflaton $   \sigma$ produces gauge
bosons: photons, gluons, $   W^{\pm}$, $Z$, and matter
fields: quarks, leptons and the Higgs bosons, while the inflaton
field $   \sigma'$ produces Sh-world particles: shadow
photons and gluons, thetons, $   W'$, $   Z'$,
theta-quarks $   q_{\theta}$, theta-leptons $  
l_{\theta}$, shadow quarks $   q'$ and leptons $ 
 l'$, scalar bosons $   \phi_{\theta}$ and shadow
Higgs fields $   \phi'$. In shadow world we end up with a
thermal bath of $   SM'$ and $   \theta$
particles. However, we assume that the density of theta-particles
is not too essential in cosmological evolution due to small $
  \theta$ coupling constants.

\end{frame}

\begin{frame}
\frametitle{Reheating and radiation}
\noindent
According to investigations of Ref. BKPRRS
\vskip 0.2in
\noindent
at the end of inflation the O- and Sh-sectors are reheated in a
non-symmetric way: $  T_R>T'_R$.
\vskip 0.2in
\noindent
After reheating (at $  T<T_R$) the exchange processes
between O- and H-worlds are too slow, by reason of very weak
interaction between two sectors. As a result, it is impossible to
establish equilibrium between them. So that both worlds evolve
adiabatically and the temperature asymmetry ($  T'/T <
1$) is approximately constant in all epochs from the end of
inflation until the present epoch.

\noindent
Therefore, the cosmology of the early Sh-world is very different
from the ordinary one when we consider such crucial epochs as
baryogenesis and nucleosynthesis.

\end{frame}

\section{Big Bang Nucleosynthesis (BBN)}

\begin{frame}
\frametitle{Big Bang Nucleosynthesis}
\noindent
At the end of cosmic inflation the Universe was filled with a
quark-gluon plasma. This plasma cools until  {\bf  the
hadron epoch} when hadrons (including baryons) can form.
\vskip 0.2in
\noindent
Then neutrinos decouple and begin travelling freely through space.
This cosmic neutrino background is analogous to the CMB which was
emitted much later.
\vskip 0.2in
\noindent
After hadron epoch the majority of hadrons and anti-hadrons
annihilate each other, leaving leptons and anti-leptons dominating
the mass of the Universe.
\vskip 0.2in
\noindent
Here we reach \alert{the lepton epoch}.

\noindent
Then the temperature of the Universe continues to fall and falls
until the stop of the lepton/anti-lepton pairs creation. Also the
most leptons/anti-leptons are eliminated by annihilation
processes.

\end{frame}

\begin{frame}
\frametitle{Big Bang Nucleosynthesis}
\noindent
At the end of the lepton epoch the Universe undergoes \alert{ 
the photon epoch} when the energy of the Universe is dominated by
photons, which still essentially interact with charged protons,
electrons and eventually nuclei.
\vskip 0.2in
\noindent
The temperature of the Universe again continues to fall. It falls
to the point when atomic nuclei begin to form. Protons and
neutrons combine into atomic nuclei by nuclear fusion process.
However, this nucleosynthesis stops at the end of the nuclear
fusion. At this time, the densities of non-relativistic matter
(atomic nuclei) and relativistic radiation (photons) are equal.

\noindent
The BBN epoch in the Sh-world proceeds differently from ordinary
one and predicts different abundances of primordial elements.

\noindent
The difference of the temperatures ($  T'<T$) gives that
the number density of H-photons is much smaller than for
O-photons:
$$   \frac {n'_{\gamma}}{n_{\gamma}}=x^3\ll 1.  $$

\end{frame}

\begin{frame}
\frametitle{Big Bang Nucleosynthesis}
\noindent
The primordial abundances of light elements depend on the baryon
to photon number density ratio: $  \eta=n_B/n_{\gamma}$.
The result of WMAP  gives: $  \eta\simeq 6\cdot
10^{-10}$, in accordance with the observational data.
\vskip 0.2in
\noindent
The universe expansion rate at the ordinary BBN epoch (with
$  T\sim 1$ MeV) is determined by the O-matter density
itself. As far as $  T'\ll T$, for the ordinary observer
it is difficult to detect the contribution of Sh-sector, which is
equivalent to $  \Delta N_{\nu}\approx 6.14x^4$ and
negligible for $  x\ll 1$.
\vskip 0.2in
\noindent
As for the BBN epoch in the shadow world, for the Sh-observer the
contribution of O-sector is equivalent to $  \Delta
N'_{\nu}\approx 6.14x^{-4}$, which is dramatically large.
Therefore, the observer in Sh-world, which measures the abundances
of shadow light elements, should immediately detect the
discrepancy between the universe expansion rate and Sh-matter
density at the shadow BBN epoch (with $  T'\sim 1 $ MeV):
the O-matter density is invisible for the Sh-observer.

\end{frame}

\begin{frame}
\frametitle{Big Bang Nucleosynthesis. Recombination}
\noindent
Then the most of electrons and protons recombine into neutral
hydrogen and free electron density strongly diminishes. During the
recombination the photon scattering rate drops below the Hubble
expansion rate.
\vskip 0.2in
\noindent
Thus, at the end of recombination, most of the atoms in the
Universe is neutral, photons travel freely and the Universe
becomes transparent. The observable CMB is a picture of the
Universe at the end of this epoch.

\end{frame}

\section{Baryon density and dark matter}

\begin{frame}
\frametitle{Baryon density and dark matter}
\noindent
Shadow baryons (and shadow helium), which are invisible by
ordinary photons, are the best candidates for dark matter (DM).
\vskip 0.2in
\noindent
Here we give an approximate estimate of baryon masses in the O-
and Sh-worlds.
\vskip 0.2in
\noindent
The most part of mass of nucleons (proton and neutron) is provided
with dynamical (constituent) quark masses $  m_q$ forming
the nucleon. The dynamical quark mass is
$$   m_q\simeq m_0 + \Lambda_{QCD},  $$
where $  m_0\sim 10\,\, {\rm{MeV}}$  is a current mass of
light quarks $  u,\,\,d$, and $$ 
\Lambda_{QCD}\simeq 300\,\,\rm{MeV}.$$
\end{frame}

\section{Baryon density and dark matter}

\begin{frame}
\frametitle{Baryon density and dark matter}
\noindent
Then the nucleon mass $  M_B$ can be estimated as
$$    M_B\simeq 3m_q\simeq 1\,\, {\rm{GeV}}.  $$
\vskip 0.2in
\noindent
As to shadow current quark mass $  m'_0$, we have
$$   m'_0\simeq \zeta m_0 \sim 1\,\, {\rm{GeV}}  $$
for $  \zeta\sim 100$.
\vskip 0.2in
\noindent
This estimate gives the shadow baryon mass $  M'_B$ equal
to
$$   M'_B\simeq 3(m'_0 + \Lambda'_{QCD}).  $$

\end{frame}

\begin{frame}
\frametitle{Baryon density and dark matter}
\noindent
Taking into account Ref.

\noindent
we obtain $$    \Lambda'_{QCD}\simeq 450\,\,\rm{MeV},$$
and:
$$     M'_B\simeq 3(1+0.45)\,\, {\rm{GeV}}\simeq
4.35\,\,{\rm{GeV}}. $$
Here we want to comment that in our model baryons of shadow world
are formed not only by quark system $    qqq$, but also
by $    q_{\theta,\vartheta}q_{\theta}^{\vartheta}q$,
where $    \vartheta=1,2$ is the index of $  
SU(2)'_\theta$-group. The last system gives the quark-diquark
structure of shadow baryons. However, they do not give  essential
contributions to baryon density, by reason of small theta-charges.

\end{frame}

\begin{frame}
\frametitle{Baryon density and dark matter}
\noindent
Since Sh-sector is cooler than the ordinary one, then we have $
   n'_B\gtrsim n_B$ by estimate of Ref. BKPRRS
\vskip 0.2in
\noindent
And:
$$   \rho'_B=n'_BM'_B > \rho_B=n_BM_B.  $$
Now we can explain the value $   \rho_{DM}$, especially
if we take into account the shadow helium mass fraction.

\end{frame}

\begin{frame}
\frametitle{Baryon density and dark matter}
\noindent
Finally, we predict that the energy density of Sh-sector is:
$$     \rho' =  \rho_{DE} + \rho_{DM} = \rho_{DE} +
\rho'_{B} + \rho_{CDM},  $$
where $   \rho'_{B}=n'_BM'_B \approx 0.17 \rho_c$ and
$   \rho_{CDM}\approx 0.04 \rho_c$ presumably contains
shadow helium.
\vskip 0.2in
\noindent
The energy density of the O-world is;
   $$    \rho_M = \rho_B + \rho_{nuclear},  $$
where $   \rho_B=n_BM_B\approx 0.04 \rho_c$ and the
contribution of ordinary helium and other atoms is much smaller.
\vskip 0.2in
\noindent
Then it is possible to explain the following observable result:
$$     \frac{\Omega_{DM}}{\Omega_{M}} \simeq \frac
{\rho_{DM}}{\rho_M}\simeq
 \frac {\rho'_{B} + \rho_{CDM}}{\rho_B + \rho_{nuclear}}\simeq \frac{0.17 +
0.04}{0.04}\simeq 5.   $$

\end{frame}

\section{Baryogenesis}

\begin{frame}
\frametitle{Baryogenesis}
\noindent
It is necessary to devote a separate talk for {\bf 
Baryogenesis}
\vskip 0.2in
\noindent
In our cosmological model with superstring-inspired $ 
E_6$ unification, the $    B-L$ asymmetry is produced by
the conversion of ordinary leptons into particles of the hidden
sector. After the non-symmetric reheating with $    T_R
> T'_R$, it is impossible to establish equilibrium
between the O- and Sh- sectors, and baryon asymmetry may be
generated even by scattering of massless particles. In our model
with $    E_6$-unification existing at the early stage of
the Universe, after the breaking of $    E_6{(E'_6)}$,
heavy Majorana neutrinos $    N_a$ become singlets of the
subgroups $    SU(3)_C\times SU(2)_L \times U(1)_X\times
U(1)_Z$ and $    SU(3)'_C\times SU(2)'_L\times U(1)'_X
\times U(1)'_Z$, and can play the role of messengers between O-
and Sh-worlds. $    B - L$ quantum number is generated in
the decays of heavy Majorana neutrinos, $    N$, into
leptons $  l$ (or anti-leptons $  \bar l$) and
the Higgs bosons $  \phi$: $  N \to
l\phi,\,\,\bar l\bar \phi$.

\end{frame}

\begin{frame}
\frametitle{Baryogenesis}
\noindent
The three necessary Sakharov conditions, given by Ref.

\noindent
are realized in our model of baryogenesis:
\begin{enumerate}
\item $  B-L$ and $  L$ are violated by the heavy
neutrino Majorana masses.
\item The out-of-equilibrium condition is satisfied due to the
delayed decay(s) of the Majorana neutrinos, when the decay rate
$  \Gamma(N)$ is smaller than the Hubble rate $ 
H$: $\Gamma(N)< H$, i.e. the life-time is larger than the age of
the Universe at the time when $  N_a$ becomes
non-relativistic.
\item CP-violation (C is trivially violated due to the chiral nature
of the fermion weak eigenstates) originates as a result of the
complex $  lN\phi$ Yukawa couplings producing asymmetric
decay rates:
$$     \Gamma(N\to l\phi)\neq  \Gamma(N\to \bar l
\bar{\phi}),  $$
so that leptons and anti-leptons are produced in different amounts
and the $  B-L$ asymmetry is generated.
\end{enumerate}
\end{frame}

\section{Conclusions}

\begin{frame}
\frametitle{Conclusions}
\begin{itemize}
\item In this paper we have developed the hypothesis of parallel
existence of the ordinary (O) and hidden (Sh) sectors of the
Universe.
\item We have constructed a new cosmological model with the
superstring-inspired $  E_6$ unification in the
4-dimensional space.
\item We have assumed that this unification was broken at the early
stage of the Universe into $$  SO(10)\times U(1)_Z$$

-- in the O-world,

and $$  SU(6)'\times SU(2)'_{\theta}$$

-- in the Sh-world.
\end{itemize}
\end{frame}

\begin{frame}
\frametitle{Conclusions}
\begin{itemize}
\item We have investigated the breaking mechanism of the $ 
E_6$ unification. In the O-world this breaking is realized with
the Higgs field $  H_{27}$ belonging to the 27-plet,
while in the hidden sector the breakdown of the $  E'_6$
unification has come true due to the Higgs field $ 
H_{351}$ belonging to the 351-plet of the $  E'_6$. The
corresponding VEVs are $  v=\langle H_{27}\rangle$ and
$  V=\langle H_{351}\rangle$.
\end{itemize}
\end{frame}

\begin{frame}
\frametitle{Conclusions}
\begin{itemize}
\item From the beginning, we have assumed that $  E'_6$ is
the mirror counterpart of the $  E_6$. Then the discrete
symmetry $  Z_2$ (connected with the mirror parity MP)
leads to the phenomenologically unacceptable wall. Using the
simplest model of inflation with the superpotential $ 
W=\lambda \varphi(\Phi^2-\mu^2)$, where the field $ 
\varphi$ is the inflaton and $  \Phi$ is the Higgs field,
($  \lambda$ is a coupling constant and $  \mu$
is a dimensional parameter of the order of the GUT scale $
 \sim 10^{18}\,\,\rm{ GeV}$), we avoid this unacceptable wall
dominance assuming the following fine-tuning:
$$  V=V',$$ what gives
$  \lambda^2\mu^4={\lambda'}^2{\mu'}^4$. Here $ 
V^{(')}={\lambda^{(')}}^2{\mu^{(')}}^4$ is the energy density of
the tree level potential.
\end{itemize}
\end{frame}

\begin{frame}
\frametitle{Conclusions}
\begin{itemize}
\item According to our assumptions, there exists the following chains
of symmetry groups:$$   E_6\to SO(10)\times U(1)_Z \to
SU(4)_C\times SU(2)_L \times SU(2)_R\times U(1)_Z$$ $$ \to
SU(3)_C\times SU(2)_L \times SU(2)_R\times U(1)_X\times
 U(1)_Z  $$ $$\to [SU(3)_C\times SU(2)_L\times U(1)_Y]_{{SUSY}}\to
 SU(3)_C\times SU(2)_L\times U(1)_Y $$
- in the O-world,

and $$   E'_6 \to SU(6)'\times SU(2)'_{\theta} \to
SU(4)'_C\times SU(2)'_L\times SU(2)'_{\theta}\times U(1)'_Z$$ $$ \to
SU(3)'_C\times SU(2)'_L\times SU(2)'_{\theta}\times U(1)'_X \times
U(1)'_Z $$ $$\to [SU(3)'_C\times SU(2)'_L\times SU(2)'_{\theta}\times
U(1)'_Y]_{{SUSY}}$$ $$\to SU(3)'_C\times SU(2)'_L\times
SU(2)'_{\theta}\times U(1)'_Y $$
- in the Sh-world.
\end{itemize}
\end{frame}

\begin{frame}
\frametitle{Conclusions}
\begin{itemize}
\item In contrast to the results of Refs.of Berezhiani at al, based
on the concept of the parallel existence in Nature of the mirror
(M-) and ordinary (O-) worlds described by a minimal symmetry
$  G_{SM}\times G'_{SM}$, we assume the existence of
low-energy symmetry group $   G' = SU(3)'_C\times
SU(2)'_L\times SU(2)'_{\theta}\times U(1)'_Y $ in the Sh-world and
the SM symmetry group in the O-world. This is a natural
consequence of different schemes of the $  E_6$-breaking
in the O- and Sh-worlds. In comparison with $  G_{SM}$,
the group $  G'$ has an additional non-Abelian $
 SU(2)'_{\theta}$ group whose gauge fields are massless vector
particles `thetons'. These `thetons' have a macroscopic
confinement radius $  1/\Lambda'_{\theta}$. The estimate
confirms the scale $  \Lambda'_{\theta} \sim 10^{-3}
\,\,\rm{eV}$.
\item Assuming the cancellation between the 'bare' cosmological
constant, $  \lambda$, and the vacuum energy stress,
$  8\pi G\rho_{vac}$, described only by the SM
contributions of the O- and Sh-worlds, we explain the small value
of $  \rho_{DE}$, i.e. the observable tiny $CC$, only as a
result of the theta-fields condensation: $  \rho_{DE} =
\rho^{(eff)}_{vac} = {(\Lambda'_\theta)}^4\simeq (2.3\times
10^{-3}\,\,{\mbox{eV}})^4$.
\end{itemize}
\end{frame}

\begin{frame}
\frametitle{Conclusions}
\begin{itemize}
\item We have discussed how the SM-particles were generated at the
end of inflation: the inflaton decays, and the subsequent
thermalization of these decay products generates the SM-particles.
The inflaton $  \sigma$ produces gauge bosons: photons,
gluons, $  W^{\pm}$, $  Z$, and matter fields:
quarks, leptons and the Higgs bosons, while the inflaton $
 \sigma'$ produces hidden particles: shadow photons, gluons and
`thetons', $  W'$, $  Z'$, theta-quarks $
 q_{\theta}$, theta-leptons $  l_{\theta}$, shadow
quarks $  q'$ and shadow leptons $  l'$, scalar
bosons $  \phi_{\theta}$ and shadow Higgs fields $
 \phi'$.
\item The O- and Sh-sectors have different cosmological evolutions:
they never had to be in equilibrium with each other. The Big Bang
Nucleosynthesis (BBN) constraints require that Sh-sector must have
smaller temperature than O-sector: $  T'<T$. The
difference between the O- and Sh-worlds is described in terms of
two macroscopic parameters: $   x\equiv {T'}/{T}, \quad
\beta\equiv {\Omega'_B}/{\Omega_B}$, where $  T(T')$ is
O-(Sh-) photon temperature of the Universe at present, and $
 \Omega_B(\Omega'_B)$ is O-(Sh-)baryons fraction.
\end{itemize}
\end{frame}

\begin{frame}
\frametitle{Conclusions}
\begin{itemize}
\item We have considered the reheating and radiation and Big Bang
Nucleosynthesis. During reheating the exponential expansion,
developed by inflation, ceases and the potential energy of the
inflaton field decays into a hot relativistic plasma of particles.
The relativistic fraction is represented by photons and neutrinos.
The radiation is dominant at the early epochs of the Universe, but
it is negligible at present epoch: $  \Omega_r^{(0)}\ll
1$.
\item The contribution of the Sh-degrees of freedom to the
observable Hubble expansion rate, which are equivalent to an
effective number of extra neutrinos $  \Delta
N_{\nu}=6.14\cdot x^4$, is small enough. In our model: $ 
\omega_r=\Omega_r h^2=4.2\cdot 10^{-5}(1+x^4)\quad (h=H/H_0)$,
where the contribution of Sh-species is negligible due to the BBN
constraint: $  x^4\ll 1$.
\end{itemize}
\end{frame}

\begin{frame}
\frametitle{Conclusions}
\begin{itemize}
\item At the end of inflation the O- and H-sectors are reheated in a
non-symmetric way: $  T_R>T'_R$. After reheating, at
$  T<T_R$, the exchange processes between O- and
Sh-worlds are too slow (by reason of very weak interaction between
two sectors), and it is difficult to establish equilibrium between
them. As a result, the temperature asymmetry ($  T'/T <
1$) is approximately constant from the end of inflation until the
present epoch.
\item We have seen that the cosmological evolutions of the early O-
and Sh-worlds are very different, in particular, when we consider
such crucial epochs as baryogenesis and nucleosynthesis. The BBN
epoch proceeds differently in the O- and Sh-worlds and predicts
different abundances of primordial elements. For example, due to
the condition $  T'<T$ the density of Sh-photons number
is much smaller than for O-photons: $ 
n'_{\gamma}/n_{\gamma}=x^3\ll 1$.
\end{itemize}
\end{frame}

\begin{frame}
\frametitle{Conclusions}
\begin{itemize}
\item The structure formation in the Universe is connected with the
plasma recombination and matter-radiation decoupling (MRD) epochs.
Also the matter-radiation equality (MRE) is important, which is
given by the relation $$   1+z_{eq} = \Omega_m/\Omega_r
\simeq 2.4\cdot 10^4\cdot\Omega_m h^2/(1+x^4).$$ During the MRD
epoch the most of electrons and protons recombine into neutral
hydrogen and the free electron density essentially diminishes. The
MRD temperature is $   T_{dec}\simeq 0.26$ eV what
corresponds to the redshift $   1+z_{dec} =
T_{dec}/T_{today}\simeq 1100$. In the Sh-world we have the MRD
temperature $   T'_{dec}\simeq T_{dec}$ and $ 
1+z'_{dec}\simeq x^{-1}(1+z_{dec})\simeq {1100}/{x}$, what means
that in the Sh-world MRD occurs earlier than in the O-world.
\end{itemize}
\end{frame}

\begin{frame}
\frametitle{Conclusions}
\begin{itemize}
\item During the recombination epoch the photon scattering rate
drops below the Hubble expansion rate. The Sh-photon decoupling
epoch coincides with the MRE epoch. At the end of recombination,
the atoms in the Universe are neutral, photons travel freely and
the Universe becomes transparent. The observation of CMB gives a
picture of the Universe at the end of this epoch.
\item We have estimated $   \rho_M$ and $ 
\rho_{DM}$ in the framework of our cosmological model. We assume
that shadow baryons and shadow helium, invisible for ordinary
photons, give the main contribution to dark matter (DM). We
explain the observable result: $   \Omega_{DM}/\Omega_{M}
\simeq \rho_{DM}/\rho_M \simeq 5$.
\end{itemize}
\end{frame}

\begin{frame}
\frametitle{Conclusions}
\begin{itemize}
\item In our cosmological model with $  E_6$ unification,
the $   B-L$ asymmetry is produced by the conversion of
ordinary leptons into particles of the hidden sector. After the
non-symmetric reheating with $   T_R
> T'_R$, it is impossible to establish equilibrium
between the O- and Sh- sectors, and baryon asymmetry may be
generated even by scattering of massless particles.

After the breaking of $   E_6{(E'_6)}$, heavy Majorana
neutrinos $   N_a$ become singlets of the subgroups
$   SU(3)_C\times SU(2)_L \times U(1)_X\times U(1)_Z$ and
$   SU(3)'_C\times SU(2)'_L\times U(1)'_X \times
U(1)'_Z$, and can play the role of messengers between O- and
Sh-worlds. $  B - L$ quantum number is generated in the
decays of heavy Majorana neutrinos into leptons or anti-leptons
and the Higgs bosons.

The three necessary Sakharov's conditions are realized in our
model of baryogenesis.
\end{itemize}
\end{frame}


\begin{thebibliography}{}
\bibitem[]{} P. Q.~Hung,
{\color{black}Nucl. Phys. B {\bf 747} (2006), 55;}
\newblock J. Phys. A {\bf 40} (2007), 6871, arXiv:0707.2791.

\bibitem[]{} C. R.~Das and L. V.~Laperashvili,
\newblock Int. J. Mod. Phys. A {\bf 23} (2008), 1863, arXiv:0712.1326 [hep-ph].

\bibitem[]{} C. R.~Das and L. V.~Laperashvili,
\newblock Phys. Atom. Nucl. {\bf 72} (2009), 377 
[Yad. Fiz.  {\bf 72} (2009), 407].
\end{thebibliography}

\begin{thebibliography}{}
\bibitem[]{} A.G.~Riess et al.,
\newblock Astron.J. {\bf 116}, 1009 (1998); ArXiv: astro-ph/9805201.

\bibitem[]{} S.J.~Perlmutter et al.,
\newblock Nature {\bf 39}, 51 (1998); Astrophys.J. {\bf 517}, 565 (1999).

\bibitem[]{} C.~Bennett et al., {\color{black}ArXiv: astro-ph/0302207.}

\bibitem[]{} D.~Spergel et al., {\color{black}ArXiv: astro-ph/0302209.}

\bibitem[]{} P.~Astier et al., {\color{black}ArXiv: astro-ph/0510447.}

\bibitem[]{} D.~Spergel et al., {\color{black}ArXiv: astro-ph/0603449.}
\end{thebibliography}

\begin{thebibliography}{}
\bibitem[]{} A.G.~Riess et al.,
\newblock Astrophys. J. Suppl. {\bf 183} (2009), 109; arXiv: 0905.0697.

\bibitem[]{} W.L. Freedman et al., 
\newblock Astrophys. J. {\bf 704} (2009), 1036; arXiv: 0907.4524.

\bibitem[]{} R. Kessler at al., 
\newblock arXiv: 0908.4274.
\end{thebibliography}

\begin{thebibliography}{}
\bibitem[]{}
P.~Astier et al., {\color{black}ArXiv: astro-ph/0510447.}
\end{thebibliography}

\begin{thebibliography}{}
\bibitem[]{} P.J.E.~Peebles and A.~Vilenkin,
{\color{black}Phys.Rev. D {\bf 59}, 063505 (1999).}

\bibitem[]{} C.~Wetterich,
{\color{black}Nucl.Phys. B {\bf 302}, 668 (1998).}

\bibitem[]{} L.J.~Hall, Y.~Nomura and S.J.~Oliver,
\newblock Phys.Rev.Lett. {\bf 95}, 141302 (2005); ArXiv: astro-ph/0503706.
\end{thebibliography}

\begin{thebibliography}{}
\bibitem[]{} D. J.~Gross, J. A.~Harvey, E.~Martinec and R.~Rohm,
\newblock Phys. Rev. Lett. {\bf 54} (1985), 502;
Nucl. Phys. B {\bf 256} (1985), 253; ibid., B {\bf 267} (1986), 75.

\bibitem[]{} P.~Candelas, G. T.~Horowitz, A.~Strominger and E.~Witten,
\newblock Nucl. Phys. B {\bf 258} (1985), 46.
\end{thebibliography}

\begin{thebibliography}{}
\bibitem[]{} M. B.~Green, J. H.~Schwarz and E.~Witten,
\newblock \emph{Superstring theory}
\newblock {\color{black}(Cambridge University Press, Cambridge, 1988)}.
\end{thebibliography}

\begin{thebibliography}{}
\bibitem[]{} P.~Athron, S.F.~King, D. J.~Miller, S.~Moretti and
R.~Nevzorov,
\newblock Phys. Rev. D {\bf 80} (2009) 035009; arXiv:0904.2169;
arXiv:0901.1192.
\end{thebibliography}

\begin{thebibliography}{}
\bibitem[]{} R.~Slansky,
\newblock {\color{black}\it Group Theory for Unified Model Building,}
Phys.Rept. {\bf 79} 1 (1981)
\end{thebibliography}

\begin{thebibliography}{}
\bibitem[]{} E. W.~Kolb, D.~Seckel and M. S. Turner,
\newblock Nature {\bf 314} (1985), 415; Fermilab-Pub-85/16-A, Jan.1985.
\end{thebibliography}

\begin{thebibliography}{}
\bibitem[]{} T.D.~Lee and C.N.~Yang, 
{\color{black}Phys. Rev. {\bf 104} (1956), 254.}
\end{thebibliography}

\vskip 0.2in
\noindent
\begin{thebibliography}{}
\bibitem[]{} I.Yu.~Kobzarev, L.B.~Okun and I.Ya.~Pomeranchuk,
\newblock Yad. Fiz. {\bf 3} (1966), 1154 [Sov. J. Nucl. Phys. {\bf 3} (1966), 837]
\end{thebibliography}

\begin{thebibliography}{}
\bibitem[]{} Z.~Berezhiani, A.~Dolgov and R. N.~Mohapatra,
\newblock Phys. Lett. B {\bf 375} (1996), 26, hep-ph/9511221.
\bibitem[]{} Z.~Berezhiani and R. N.~Mohapatra,
\newblock Phys. Rev. D {\bf 52} (1995), 6607, hep-ph/9505385.
\bibitem[]{} Z.~Berezhiani,
{\color{black}{\it Through the looking-glass: 
Alice's adventures in mirror world},
in: Ian Kogan Memorial Collection ``From Fields to Strings: 
Circumnavigating Theoretical Physics'',
Ed. M.~Shifman et.~al., World Scientific, Singapore,
Vol.~3, pp. 2147-2195, 2005, hep-ph/0508233.}
\end{thebibliography}

\begin{thebibliography}{}
\bibitem[]{} R.~Foot, H.~Lew and R.R.~Volkas,
\newblock Phys.Lett.B {\bf 272} (1991), 67; 
Mod.Phys.Lett.A {\bf 7} (1992), 2567.
\bibitem[]{} R.~Foot,
\newblock Mod.Phys.Lett.A {\bf 9} (1994), 169.
\bibitem[]{} R.~Foot and R.R.~Volkas,
\newblock Phys.Rev.D {\bf 55} (1995), 5147.
\bibitem[]{} Review by R.~Foot
\newblock Int.J.Mod.Phys.D {\bf 13} (2004), 2161.
\end{thebibliography}

\begin{thebibliography}{}
\bibitem[]{} R.~Slansky,
\color{black}{\it Group Theory for Unified Model Building,}
\newblock Phys.Rept. {\bf 79} 1 (1981))
\end{thebibliography}

\begin{thebibliography}{}
\bibitem[]{} C.R.~Das, L.V.~Laperashvili, H.B.~Nielsen and A.~Tureanu,
\newblock\color{black}{\it Mirror World and Superstring-Inspired Hidden Sector
of the Universe, Dark Matter and Dark Energy.}
\newblock \color{black}{arXiv:1101.4558 [hep-ph], to appear in Phys.Rev.D (2011).}
\end{thebibliography}

\begin{thebibliography}{}
\bibitem[]{} Taichiro Kugo, Joe Sato,
\newblock Prog. Theor. Phys. {\bf 91}
(1994) 1217; hep-ph/9402357.
\end{thebibliography}

\begin{thebibliography}{}
\bibitem[]{} P.Q.~Hung,
\newblock Nucl. Phys. B {\bf 747} (2006) 55; J.
Phys. A {\bf 40} (2007) 6871;
\bibitem[]{} P.Q.~Hung and P.~Mosconi, {\color{black}hep-ph/0611001;}
\bibitem[]{} M.~Adibzadeh and P. Q.~Hung, {\color{black}Nucl. Phys. B {\bf 804} (2008) 223;}
\bibitem[]{} H.~Goldberg, {\color{black}Phys. Lett. B {\bf 492} (2000) 153;}
\bibitem[]{} C. R.~Das and L.V.~Laperashvili, 
\newblock Int. J. Mod. Phys. A {\bf 23} (2008) 1863;
arXiv:0712.1326 [hep-ph;  Phys. Atom. Nucl. {\bf 72} (2009) 377;
arXiv:0712.0253 [hep-ph].
\end{thebibliography}

\begin{thebibliography}{}
\bibitem[]{} C.R.~Das, L.V.~Laperashvili and A.~Tureanu,
\newblock Eur.Phys.J.C
{\bf 66} (2010) 307; arXiv:0902.4874; AIP Conf.Proc. {\bf 1241}
(2010) 639; arXiv:0910.1669.
\end{thebibliography}

\begin{thebibliography}{}
\bibitem[]{} G.~Dvali, Q.~Shafi, R.~Schaefer, \color{black}{Phys.Rev.Lett. {\bf 73} (1994)
1886.}
\end{thebibliography}

\begin{thebibliography}{}
\bibitem[]{} Z.~Berezhiani, D.~Comelli and N.~Tetradis, 
\newblock Phys. Lett. B {\bf 431} (1998) 286.
\end{thebibliography}

\begin{thebibliography}{}
\bibitem[]{} Y. B.~Zeldovich, \color{black}{JETP Lett. {\bf 6} 316 (1967).}
\bibitem[]{} S.~Weinberg, \color{black}{Rev. Mod. Phys. {\bf 61} 1 (1989).}
\end{thebibliography}

\begin{thebibliography}{}
\bibitem[]{} M. T.~Veltman, \color{black}{Phys. Rev. Lett. {\bf 34} (1975) 777.}
\end{thebibliography}

\begin{thebibliography}{}
\bibitem[]{} C. D.~ Froggatt, L. V.~Laperashvili, R. B.~Nevzorov
and H. B.~Nielsen, 
\newblock Phys. Atom. Nucl. {\bf 67} (2004), 582 [Yad.
Fiz. {\bf 67} (2004), 601]; arXiv:hep-ph/0310127; Proceedings of
7th Workshop on 'What Comes Beyond the Standard Model', Bled,
Slovenia, 19-30 Jul 2004; published in *Bled 2004, What comes
beyond the standard models*, pp. 17-27, DMFA-Zaloznistvo,
Ljubljana, 2004; hep-ph/0412208, hep-ph/0411273.
\bibitem[]{} C.~Froggatt, R.~Nevzorov and H. B. Nielsen, 
\newblock Nucl. Phys. B {\bf 743} (2006) 133, hep-ph/0511259; 
J. Phys. Conf. Ser. {\bf 110}
(2008) 072012; arXiv:0708.2907 [hep-ph].
\end{thebibliography}

\begin{thebibliography}{}
\bibitem[]{} D. L.~Bennett and H. B.~Nielsen,\newblock Int. J. Mod. Phys. A
{\bf 9} (1994), 5155; ibid., A {\bf 14} (1999) 3313;
\bibitem[]{}  C. D.
Froggatt and H. B. Nielsen,\newblock {\it Origin of Symmetries} (World
Scientific, Singapore, 1991);
\bibitem[]{}C. D. Froggatt and H. B. Nielsen,\newblock
Phys. Lett. B {\bf 368} (1996) 96.
\end{thebibliography}

\begin{thebibliography}{}
\bibitem[]{} D.J.~Shaw, J.D.~Barrow, 
\newblock {\it A Testable Solution of
the $CC$ and Coincedence Problems}, arXiv:1010.4262[gr-qc].
\end{thebibliography}

\begin{thebibliography}{}
\bibitem[]{} E.~Baum, \color{black}{Phys. Lett. B {\bf 133} (1983) 185,}
\bibitem[]{} S.~Hawking, \color{black}{Phys. Lett. B {\bf 134} (1984) 403.}
\end{thebibliography}

\begin{thebibliography}{}
\bibitem[]{}  Z.~Berezhiani,\newblock  {\it Through the looking-glass:
Alice's adventures in mirror world}, in: Ian Kogan Memorial
Collection ``From Fields to Strings: Circumnavigating Theoretical
Physics'', Eds. M.~Shifman et al., World Scientific, Singapore,
Vol.~3, pp. 2147-2195, 2005;
 AIP Conf. Proc. {\bf 878} (2006)
195; Eur. Phys. J. ST {\bf 163} (2008) 271.
\end{thebibliography}

\begin{thebibliography}{}
\bibitem[]{} Z.~Berezhiani, L.~ Kaufmann, P.~Panci, N.~Rossi,
A.~Rubbia and A.~Sakharov, 
\newblock {\it Strongly interacting mirror dark
matter}, CERN-PH-TH-2008-108, May 2008.
\end{thebibliography}

\begin{thebibliography}{}
\bibitem[]{} Z.~Berezhiani, 
\newblock {\it Through the looking-glass: Alice's
adventures in mirror world}, in: Ian Kogan Memorial Collection
``From Fields to Strings: Circumnavigating Theoretical Physics'',
Eds. M.~Shifman et al., World Scientific, Singapore, Vol.~3, pp.
2147-2195, 2005; AIP Conf. Proc. {\bf 878} (2006) 195; Eur. Phys.
J. ST {\bf 163} (2008) 271
\end{thebibliography}

\begin{thebibliography}{}
\bibitem[]{}  A.D.~Sakharov,  \color{black}{Pisma Zh. Eksp. Teor. Fiz. {\bf 5}
(1967) 32.}
\end{thebibliography}
\end{document}